%% file: main.tex
\documentclass[accepted]{melba}

\usepackage{mwe} 

\usepackage{amsmath,amsfonts}
\usepackage{url}            
\usepackage{booktabs,makecell,fvextra}       
\usepackage{amsfonts}       
\usepackage{nicefrac}       
\usepackage{microtype}      
\usepackage{xcolor}         
\usepackage{amsmath,amssymb,graphicx,amsthm,cleveref}
\usepackage{enumitem,listings}
\usepackage{inconsolata}
\usepackage[T1]{fontenc}
\usepackage{orcidlink,textcomp}
\usepackage{float,siunitx,caption}
\usepackage{multirow,diagbox,calc,cuted}
\usepackage{tcolorbox}
\tcbuselibrary{skins}
\input{defs}

\melbaid{2026:028}  
\doi{10.59275/j.melba.2026-b6a3}
\melbaauthors{Wang, McDonagh and Davies}
\email{andrew.wang@ed.ac.uk}
\volume{2026}
\firstpageno{571}  
\melbayear{2026}  
\datesubmitted{2026-02-22}  
\datepublished{2026-08-29}  

\ShortHeadings{Benchmarking Self-Supervised Learning Methods for Accelerated MRI Reconstruction}{Wang, McDonagh and Davies}

\title{Benchmarking Self-Supervised Learning Methods\\for Accelerated MRI Reconstruction}

\author{
	\name Andrew Wang\aff{1}\orcidlink{0000-0003-0838-7986},
	\name Steven McDonagh\aff{1}\orcidlink{0000-0001-7025-5197},
    \name Mike Davies\aff{1}\orcidlink{0000-0003-2327-236X}
}
\affiliations{%
	\num 1 \addr Institute for Imaging, Data and Communications, School of Engineering, University of Edinburgh
}

\abstract{%

\input{sec/0_abstract}
}

\keywords{self-supervised learning, image reconstruction, MRI, inverse problems, benchmarking}

\begin{document}
\twocolumn[\maketitle]

\input{sec/1_intro}
\input{sec/2_ssibench}
\input{sec/3_experiments}
\input{sec/4_discussion}

\acks{This work was supported by the School of Engineering at the University of Edinburgh.}

\ethics{This study was conducted retrospectively using human subject data made available in open access by \citet{zbontar_fastmri_2018,desai_skm-tea_2022,wang_cmrxrecon_2023,yu_validation_2022}. Ethical approval was not required as confirmed by the licenses attached with the open access data.}

\coi{The authors have no relevant financial or non-financial interests to disclose.}

\data{The data supporting the findings of this study are openly available from \citet{zbontar_fastmri_2018,desai_skm-tea_2022,wang_cmrxrecon_2023,yu_validation_2022}, subject to their respective licenses. The code is available at \url{https://github.com/Andrewwango/ssibench} and in DeepInverse \citep{tachella_deepinverse_2025} at \url{http://deepinv.org}.}
\bibliography{references}

\input{sec/X_suppl}

\end{document}

%% file: defs.tex
\usepackage[textsize=tiny]{todonotes}

\def\x{{\mathbf x}}

\def\xhat{{\mathbf{\hat x}}}
\def\y{{\mathbf y}}
\def\A{{\mathbf A}}
\def\AT{{\mathbf{A}^{\top}}}
\def\M{{\mathbf M}}
\def\T{{\mathbf T}}
\def\Tg{{\mathbf{T}_g}}

\def\L{{\mathcal{L}}}

\def\ft{{f_\theta}}
\newcommand{\std}[1]{\textsubscript{\textpm{}#1}}
\newcommand{\sstd}[1]{\textsuperscript{\makebox[0pt][l]{*}}\std{#1}}

\definecolor{codegreen}{rgb}{0,0.6,0}
\definecolor{codegray}{rgb}{0.5,0.5,0.5}
\definecolor{codepurple}{rgb}{0.58,0,0.82}
\definecolor{backcolour}{rgb}{0.95,0.95,0.92}
\lstdefinestyle{mystyle}{
    commentstyle=\color{codegreen},
    keywordstyle=\color{magenta},
    numberstyle=\tiny\color{codegray},
    stringstyle=\color{codepurple},
    basicstyle=\ttfamily\scriptsize,
}
\newtheorem*{remark}{Remark}
\theoremstyle{definition}
\newtheorem*{proposition}{Proposition}

\RequirePackage{xspace}
\makeatletter
\DeclareRobustCommand\onedot{\futurelet\@let@token\@onedot}
\def\@onedot{\ifx\@let@token.\else.\null\fi\xspace}

\def\eg{\emph{e.g}\onedot} 

\def\ie{\emph{i.e}\onedot}

\makeatother

\newtcolorbox{rebuttalbox}{
  enhanced,
  colframe=yellow,
  colback=white,
  coltitle=black,
  colbacktitle=yellow,
  title=NEW,
  attach boxed title to top left,
  boxed title style={size=small}
}

%% file: sec/0_abstract.tex
Reconstructing MRI from highly undersampled measurements is crucial for accelerating medical imaging, but is challenging due to the ill-posedness of the inverse problem. While supervised deep learning (DL) approaches have shown remarkable success, they traditionally rely on fully-sampled ground truth (GT) images, which are expensive or impossible to obtain in real scenarios. This problem has created a recent surge in interest in self-supervised learning methods that do not require GT. Although recent methods are now fast approaching ``oracle'' supervised performance, the lack of systematic comparison and standard experimental setups are hindering targeted methodological research and precluding widespread trustworthy industry adoption. We present \mbox{\textbf{SSIBench}}, a modular and flexible comparison framework to unify and thoroughly benchmark \textbf{S}elf-\textbf{S}upervised \textbf{I}maging methods (SSI) without GT. We focus on end-to-end trained DL methods, which do not require long inference time, large datasets, or per-image training. We evaluate 21 such recent methods across seven realistic MRI scenarios on real data, showing a wide performance landscape whose method ranking differs across scenarios and metrics, exposing the need for further SSI research.
To accelerate reproducible research and lower the barrier to entry, we provide the extensible benchmark and open-source reimplementations of all methods at \href{https://github.com/Andrewwango/ssibench}{https://github.com/Andrewwango/ssibench}, 
allowing researchers to rapidly and fairly contribute and evaluate new methods on the standardised setup for potential leaderboard ranking, or benchmark existing methods on custom datasets, forward operators, or models, unlocking the application of SSI to other valuable nascent GT-free scientific imaging modalities.

%% file: sec/1_intro.tex
\section{Introduction}
\label{sec:intro}

\begin{figure*}
\centering
\includegraphics[width=0.95\linewidth]{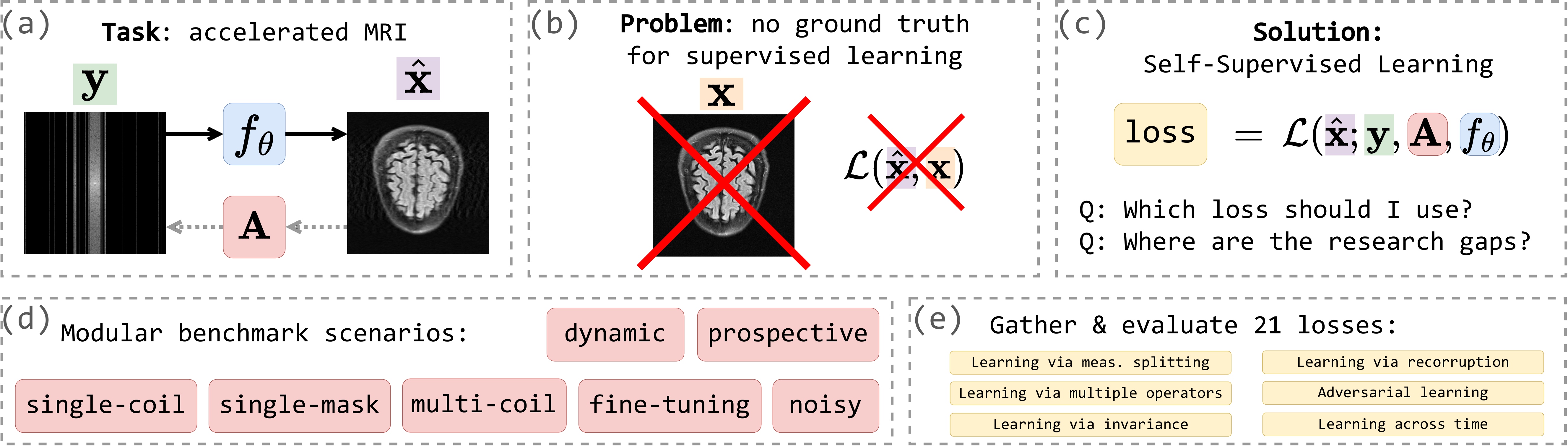}
\caption{\textit{(a)} Accelerated MRI reconstruction from undersampled images $\y$. \textit{(b)} Ground truth (GT) images $\x$ are impossible or expensive to obtain. \textit{(c)} There has been a recent surge in self-supervised imaging (SSI) methods that can learn without GT. \textit{(d)} We introduce \mbox{\textbf{SSIBench}}, a modular benchmark over seven diverse MRI acquisition settings and anatomies, that gathers and evaluates 21 state-of-the-art end-to-end training loss functions \textit{(e)}.}
\label{fig:ssibench}
\end{figure*}

Accelerating MRI reconstruction is crucial for reducing lengthy scan times, modelled as \citep{lustig_compressed_2008}:

\begin{equation}
\y_{i,c}=\A_{i,c}\x_i+\epsilon,\;\A_{i,c}=\M_i\mathbf{F}\mathbf{S}_c,\;\epsilon\sim\mathcal{N}(\mathbf{0},\sigma^2\mathbf{I})
    \label{eq:mri_problem},
\end{equation}

where $\x_i\in\mathcal{X}\in\mathbb{R}^n$ is the $i$-th underlying image, $\y_i=\{\y_{i,c}\}_{c=1}^C\in\mathcal{Y}\in\mathbb{R}^m$ are the \textit{$k$-space} measurements, $\A_i=\{\A_{i,c}\}_{c=1}^C$ is the \textit{forward operator} (or \textit{acquisition method}), consisting of the undersampling mask $\M_i$, the $c$-th coil sensitivity map $\mathbf{S}_c$ and the Fourier operator $\mathbf{F}$, and $m=n/\alpha$ where $\alpha$ is the acceleration rate. Note we may drop the index $i$ for clarity where it is unambiguous. The goal is to recover high quality images $\xhat_i=\ft(\y_i,\A_i)$ via a \textit{model} $\ft$ from partial, noisy measurements. However, this is challenging as the inverse problem is ill-posed, either due to undersampling $C=1,m<n$, leading to a non-trivial null-space $\mathcal{N}(\A)\in\mathbb{R}^{n-m}$, or poor conditioning $\kappa(\A)\gg1$.

Classical compressed sensing solves the problem using regularised optimisation \citep{lustig_compressed_2008}. However, this requires hand-crafted, image-specific priors, inference is lengthy and expensive, and acceleration rates may be limited \citep{heckel_deep_2024}. 
Deep learning has been used to learn $\ft$ directly from data, and achieves impressive results compared to classical methods \citep{heckel_deep_2024,klug_analyzing_2023,tachella_unsupervised_2022}. Prevailing methods assume that underlying ground-truth (GT) images $\x$ are available in order to train with supervised learning $\L_\text{sup}=\ell(\xhat,\x)$ where $\ell(\cdot,\cdot)$ is any metric such as the mean squared error (MSE). 

However, fully-sampled GT measurements are impossible or expensive to obtain in real scenarios, \eg when imaging moving organs \citep{wang_fully_2024}, adaptation to new anatomies \citep{darestani_test-time_2022}, 4D MRI \citep{ong_extreme_2020}, low-field MRI \citep{janjusevic_self-supervised_2025}, or where it leads to blurring in patient motion. Therefore, \textbf{self-supervised imaging (SSI) \ie methods that learn without access to GT} $\x$ are needed. Note this differs from ``self-supervision'' in representation learning: here, images are output $\xhat=\ft(\y,\A)$ rather than input, and GT $\x$ refers to images rather than labels. 

A vast array of SSI methods have been proposed in recent years for MRI reconstruction. However, they often fail to compare to existing methods \citep{cole_fast_2021,chen_robust_2022,millard_clean_2024}, exacerbated by disjoint communities of researchers working on the same problem, \eg from ML research, imaging, or various medical or other application domains. It is hence difficult to draw direct comparisons due to different or proprietary setups, datasets, a vast variety of models $\ft$, forward operators $\A$, evaluation protocols or lack of transparent implementation of compared methods. Typically, each proposed method claims to perform best, preventing research from focussing on where true gaps lie, and impeding trustworthy translation of SSI into real-world settings.

Our benchmark unifies previously fractured efforts by providing a standardised, objective evaluation of methods, leading to trustworthy application. It also provides the opportunity for ML researchers to find unsolved gaps exposed in the common framework and clear performance landscape, and collectively contribute methodological advances to solve real-world and future imaging problems. By making the benchmark modular and accessible to the public, researchers can \textit{a)} rapidly implement new methods or combine existing ones and fairly test these on the standard setup, \textit{b)} evaluate them on custom datasets, models $\ft$, or forward operators $\A$, or \textit{c)} use as a blueprint for other challenging nascent scientific imaging problems within and beyond MRI, where lack-of-GT currently hurts progress.

Our benchmark investigates methods that satisfy critical inclusion criteria for practical adoption, such as speed, ease of training and generalisability \citep{heckel_deep_2024}. Explicitly, we include:

\textbf{End-to-end trained methods}, since they provide fast inference by requiring only one neural function evaluation per image (NFE) $N=1$, and can be easily trained/fine-tuned on an modest size dataset (${\sim}500$ images) \citep{heckel_deep_2024} via feedforward inference on a discriminative model. We exclude generative methods with lengthy inference-time iterative procedures \citep{shafique_mri_2024} as these are infeasibly expensive in real-world clinical workflows, and that require large-scale datasets (${\sim}10^4$), since these are challenging to gather for many anatomies/acquisitions (such as 3D) and impede fine-tuning to new acquisitions. These include diffusion models \citep{daras_ambient_2023,kawar_gsure-based_2024,aali_ambient_2024} ($N{\sim}10^{3 \text{ to }4}$), unconditional GANs \citep{bora_ambientgan_2018} ($N{\sim}10^{2\text{ to }3}$), or flow matching \citep{luo_unsupervised_2025} ($N{\sim}10^{1\text{ to }2}$). We do include pilot results on SotA diffusion models pretrained on much larger quantities of measurements in \cref{sec:diffusion}, since these cannot be successfully trained on any of the scenarios' modest-sized datasets we consider here.
We also exclude single-image methods that require retraining for \textit{every} image, such as Deep Image Prior methods \citep{darestani_accelerated_2021,korkmaz_unsupervised_2022,zou_dynamic_2021} ($N{\sim}10^{3\text{ to }4}$), implicit neural representations \citep{shen_nerp_2024} ($N=10^3$) or test-time training \citep{qin_ground-truth_2023,darestani_test-time_2022,joo_aespa_2025} ($N{\sim}10^2$).
    
\textbf{Ground-truth-free} \ie self-supervised methods that learn from $k$-space measurements only $\{\y_i\}$, as GT images $\{\x_i\}$ are expensive or impossible to obtain during training/pretraining of models, priors or features \citep{daras_survey_2024,chung_score-based_2022,ericsson_self-supervised_2022,he_comparative_2020}. Note that GT $\{\x_i\}$ is also required for learning from unpaired data $\{\x_i\},\{\y_j\}$ \citep{oh_unpaired_2020,li_progressive_2024,lei_wasserstein_2021,mukherjee_end--end_2021} or weakly/semi-supervised learning $\{\x_i,\y_i\},\{\y_j\}$ \citep{desai_vortex_2022}.
    
\textbf{Architecture-agnostic methods} that do not depend on a specific model architecture, such that the learning is primarily guided by the loss function rather than the inductive bias of strong, specific architectural priors \citep{darestani_accelerated_2021,korkmaz_unsupervised_2022,tamir_unsupervised_2020}, and the loss function is the conceptual advance \citep{chen_equivariant_2021,millard_theoretical_2023}. This way, our findings generalise to diverse future architectures, rather than limiting the applicability of our benchmark to current modality-specific architectures.

We provide extensive experimental results benchmarking these methods on accelerated MRI reconstruction without GT across seven acquisition scenarios \& anatomies that are sufficiently general to be applied to future problems, and provide the loss reimplementations, living benchmark and training code at \href{https://github.com/Andrewwango/ssibench}{https://github.com/Andrewwango/ssibench}.

\subsection{Contributions}

\begin{enumerate}
    \item A comprehensive review unifying SotA self-supervised end-to-end trained deep learning methods for imaging;
    \item Reproducible, well-documented and \verb|pytest|ed reimplementations of 21 methods, and a modular benchmark site facilitating contribution and application to new problems;
    \item Benchmarking experiments for accelerated MRI without GT across seven insightful scenarios on a standardised setup, where different methods have different strengths, and a decision framework for practitioners.
\end{enumerate}

%% file: sec/2_ssibench.tex
\section{SSIBench - a benchmark for self-supervised imaging}
\subsection{Benchmark methodology}

We provide a systematic review of existing SotA self-supervised end-to-end deep learning methods from the literature that fall under our scope mentioned above, and unify these by constructing \textbf{SSIBench} as follows. It is the design of the loss function $\L(\xhat;\y,\A,\ft)$ that lets a method learn to recover information from the null-space \citep{tachella_self-supervised_2026}, and is the key conceptual advance that differs between methods \citep{millard_theoretical_2023,tachella_unsupervised_2022}. To provide a fair, substantive comparison and maximise compatibility of our benchmark to different setups, we evaluate the losses while fixing all other elements of the experimental setup, including the forward operator $\A$, the type, architecture and size of the model $\ft$ (such that all methods have the same inductive bias), data preprocessing stages, and the metrics. This decision is crucial for controlled comparisons of loss functions, so that the modular benchmark can easily be adapted with any model $\ft$, forward operator $\A$ or dataset of the future.

In this framework, we gather 21 distinct ground-truth-free loss functions $\L(\xhat;\y,\A,\ft)$ from across the ML and imaging literatures, summarised in \cref{tab:methods}. We reimplement these since the compared methods' original codebases often adhere to differing software engineering principles and are often implemented for their specific experimental setups with poor generalisability to other setups. We provide details of their generalised implementations and sketches of the code mapped to their equations in \cref{sec:loss_implementations}. We contribute the losses into the DeepInverse library \citep{tachella_deepinverse_2025} following standard modern coding best practice of unit-testing, code-review and thorough documentation.

\begin{table*}[t]
\centering
\caption{Self-supervised losses gathered, reimplemented and evaluated in our benchmark framework. Above: losses for reconstruction. Below: losses for joint reconstruction and denoising. NFE: number of neural function $\ft$ evaluations during loss forward pass / inference.
}
\label{tab:methods}
\resizebox{\textwidth}{!}{%
\begin{tabular}{lllll}
\hline
Method & Ref & Loss & Code: \texttt{deepinv.loss.}\rule{1cm}{0.25mm} & NFEs {\footnotesize(train/test)} \\ \hline
MC & \citep{senouf_self-supervised_2019} & \cref{eq:mc} & \texttt{MCLoss()} & $1/1$ \\
SSDU & \citep{yaman_multi-mask_2022} & \cref{eq:splitting} & \texttt{SplittingLoss()} & $1/1$ \\
Noise2Inverse & \citep{hendriksen_noise2inverse_2020} & \cref{eq:n2i} & \texttt{SplittingLoss()} & $1/J$ \\
Weighted-SSDU & \citep{millard_theoretical_2023} & \cref{eq:millard} & \texttt{WeightedSplittingLoss()} & $1/1$ \\
SSDU-Consistency & \citep{hu_self-supervised_2021} & \cref{eq:ssdu_consistency} & \texttt{SplittingConsistencyLoss()} & $2/1$ \\
MOC-SSDU & \citep{zhang_cycle-consistent_2024} & \cref{eq:moc}$+$\cref{eq:splitting} & \texttt{MOConsistencyLoss() + SplittingLoss()} & $2/1$ \\
Adversarial & \citep{cole_fast_2021} & \cref{eq:cole} & \texttt{UnsupAdversarialGeneratorLoss()} & $1/1$ \\
UAIR & \citep{pajot_unsupervised_2018} & \cref{eq:uair} & \texttt{UAIRGeneratorLoss()} & $2/1$ \\
VORTEX & \citep{desai_vortex_2022} & \cref{eq:vortex} & \texttt{AugmentConsistencyLoss()} & $2/1$ \\
ES & \citep{sechaud_equivariant_2025} & \cref{eq:es} & \texttt{ESLoss()} & $1/1$ \\
E2I & \citep{schut_equivariance2inverse_2025} & \cref{eq:splitting}+\cref{eq:ei} & \texttt{SplittingLoss() + EILoss()} & $2/1$ \\
EI & \citep{chen_equivariant_2021} & \cref{eq:ei} & \texttt{EILoss()} & $2/1$ \\
MOI & \citep{tachella_unsupervised_2022} & \cref{eq:moi} & \texttt{MOILoss()} & $2/1$ \\
MO-EI & This work & \cref{eq:moei} & \texttt{MOEILoss()} & $2/1$ \\ \hline \noalign{\vspace{0.7mm}}
\noindent ENSURE & \citep{aggarwal_ensure_2023} & \cref{eq:ensure} & \texttt{ENSURELoss()} & $2/1$ \\
DDSSL & \citep{quan_dual-domain_2022} & \cref{eq:r2r}$+$\cref{eq:ei} & \texttt{R2RLoss() + EILoss()} & $2/J$ \\
Robust-SSDU & \citep{millard_clean_2024} & \cref{eq:robust_ssdu} & \texttt{RobustSplittingLoss()} & $1/1$ \\
Noise2Recon-SSDU & \citep{desai_noise2recon_2022} & \cref{eq:millard}$+$\cref{eq:vortex} & \makecell[l]{\texttt{WeightedSplittingLoss()}\\\texttt{   + AugmentConsistencyLoss()}} & $2/1$ \\
ES-R2R & \citep{sechaud_equivariant_2025} & \cref{eq:r2r} \& \cref{eq:es} & \texttt{R2RLoss()} \& \texttt{ESLoss()} & $1/1$ \\
Robust-EI & \citep{chen_robust_2022} & \cref{eq:ei}$+$\cref{eq:sure} & \texttt{EILoss() + SUREGaussianLoss()} & $3/1$ \\
Robust-MO-EI & This work & \cref{eq:moei}$+$\cref{eq:sure} & \texttt{MOEILoss() + SUREGaussianLoss()} & $3/1$ \\ \hline
\end{tabular}%
}
\end{table*}

\subsection{Benchmark scenarios}

We consider seven common accelerated MRI experimental scenarios. This provides insight on the performance of the methods on (a) individual controlled tasks, which permit disentangling true reconstruction ability from other MRI sub-tasks \eg coil sensitivity map estimation \citep{millard_theoretical_2023,yaman_self-supervised_2020}, and (b) diverse GT-free acquisition scenarios in real-world practice \citep{yu_validation_2022,wang_cmrxrecon_2023,desai_skm-tea_2022}.

\begin{enumerate}[align=left, label=\textbf{Scen. \arabic*}, leftmargin=*]        
    \item(Single-coil) A basic but very important setup where no additional information is provided by multiple coils $C=1$, so we can demonstrate performance on recovering data from the clear null-space of the rank-deficient forward operator $\mathcal{N}(A)\in\mathbb{R}^{n-m}$. No added noise $\sigma=0$ and retrospectively undersampled to $6\times$ acceleration, with a random mask per sample $\M_i\sim\mathcal{M}$. 
    \item(Noisy) Like scenario 1 (same $\A$, data), but simulates more realistic acquisition under thermal device noise $\sigma=0.1$ \citep{millard_clean_2024}, where methods must perform joint reconstruction and denoising.
    \item(Single-mask) Like scenario 1, but with one fixed sampling mask $\M_i=\M\,\forall i$, common in clinical systems where a predetermined vendor-optimized sampling pattern is used to simplify hardware constraints and ensure consistency across patients. 
    \item(Multi-coil) Like scenario 1, but $C=4$ as in modern parallel imaging \citep{sriram_end--end_2020}. Since now $m$ is a factor of $C\times$ larger, the effective null-space is smaller; see \cref{sec:svd} for an analysis. We assume known sensitivity maps to disentangle the estimation problem.
    \item(Fine-tuning) We test the widespread real-world practice of fine-tuning a foundation model \citep{terris_reconstruct_2025} on out-of-domain data without GT on raw multicoil SKM-TEA \citep{desai_skm-tea_2022} knees at extreme 32$\times$ acc. disk sampling with estimated coil maps.
    \item(Dynamic) Learning to reconstruct 2D+t dynamic cine $10\times$ acquisition without GT on the real CMRxRecon \citep{wang_cmrxrecon_2023} cardiac dataset, as true GT is impossible to capture in dynamic MRI \citep{wang_fully_2024}.
    \item(Prospective) Learning to reconstruct a single knee volume from 7$\times$ prospectively undersampled raw data \citep{yu_validation_2022}, where fully-sampled GT truly does not exist.
\end{enumerate}

\subsection{Benchmarked methods}
\label{sec:benchmark}

We formulate the 21 loss functions gathered from across the literature; reimplementation details are in \cref{sec:loss_implementations}.

\vspace{0.5em}\noindent\textbf{Measurement consistency (MC)}\quad
The most basic self-supervised loss \citep{senouf_self-supervised_2019} simply computes:

\begin{equation}
    \L_\text{MC}=\ell(\A\xhat,\y)
    \label{eq:mc},
\end{equation}

The MC loss cannot recover information from the operator's null-space $\mathcal{N}(\A)$ as any solutions $\A^\dagger\y+\mathbf{v}$ trivially satisfy $\L_\text{MC}=0$ where $\mathbf{v}\in\mathcal{N}(\A)$. Instead, MC has been used in the presence of strong inductive bias where regularisation is provided by very specific architectures \citep{darestani_accelerated_2021,korkmaz_unsupervised_2022,tamir_unsupervised_2020} or initialisations \citep{darestani_test-time_2022}. However, to disentangle this we focus on comparing the performance of different loss functions while using identical networks: this is a common setting \citep{yaman_zero-shot_2022} where MC \citep{senouf_self-supervised_2019} simply recovers $\A^\dagger\y$ \citep{pruessmann_sense_1999}.
Traditional regularising terms (\eg sparsity) \citep{wang_neural_2020,alcalar_convex_2024} can always be appended independently to any of the losses used in this paper such as MC, so we do not consider their usage here.

\vspace{0.5em}\noindent\textbf{Learning from measurement splitting}\quad
Methods such as SSDU \citep{yaman_self-supervised_2020} and similar works \citep{yaman_multi-mask_2022,huang_self-supervised_2024,klug_analyzing_2023}, randomly split $\y$ into two sets at each instance during training:
\begin{equation}
    \L_\text{SSDU}=\ell(\M^{(2)}\A \ft(\M^{(1)}\y,\M^{(1)}\A),\M^{(2)}\y),\;\xhat=\ft(\y,\A)
    \label{eq:splitting}
\end{equation}

\noindent where $\M^{(1)}$ is a randomly generated mask (spanning the whole measurement domain in expectation), $\M^{(1)}+\M^{(2)}=\mathbb{I}_m$, and $\ell$ is either a metric in the measurement domain and image domain. Methods requiring pairs of measurements of the same subject \citep{xia_training_2019,hu_spicer_2024,gan_deep_2021,huang_deep_2019,gan_deformation-compensated_2022,liu_rare_2020} are equivalent to SSDU but with $2\times$ less acceleration. The same concept can be applied to a specific acquisition \citep{wang_k-band_2024}. However, for the estimator $\xhat$ to correspond to the supervised MMSE, an inference-time adaptation must be made to average over $J>1$ passes, which was used in both of \citet{hendriksen_noise2inverse_2020,gruber_sparse2inverse_2024}:
\begin{equation}
    \xhat=\frac{1}{J}\Sigma_{j=1}^J \ft(\M^{(j)}\y,\M^{(j)}\A)
    \label{eq:n2i}.
\end{equation}

\citet{millard_theoretical_2023} bypass this averaging by weighting the SSDU loss metric $\ell(\cdot,\cdot)$ to directly recover the MMSE estimator, and by splitting $\M^{(j)}\sim \mathcal{M}$, where $\mathbf{P}=\mathbb{E}[\M],\tilde{\mathbf{P}}=\mathbb{E}[\M^{(j)}]$ \ie the expectations of the imaging mask and the splitting mask, respectively:

\begin{equation}
\begin{split}
\L_\text{Weighted-SSDU} &= \L_\text{SSDU},\quad 
\ell(a,b):=\lVert(1-\mathbf{K})^{-\frac{1}{2}}(a-b)\rVert_2^2,\\[-2pt]
\mathbf{K} &= (\mathbb{I}_m-\tilde{\mathbf{P}}\mathbf{P})^{-1}(\mathbb{I}_m-\mathbf{P}).
\end{split}
\label{eq:millard}
\end{equation}

\citep{hu_self-supervised_2021,zhou_dual-domain_2022,wang_parcel_2023,xu_self-supervised_2025} split $\y$ with both $\M^{(1)},\M^{(2)}$, compute a SSDU-style loss for each of these subsets, and also compute a consistency loss between the two reconstructions $\xhat^{(1)},\xhat^{(2)}$, where $\bar\A=(\mathbb{I}_m-\M)\mathbf{F}\mathbf{S}$:

\begin{equation}
\begin{split}
\L_\text{SSDU-Consistency} &= 
\ell(\A\xhat^{(1)},\y)+\ell(\A\xhat^{(2)},\y)\\
&+ \ell(\bar\A\xhat^{(1)},\bar\A\xhat^{(2)}),\\[-2pt]
\xhat^{(k)} &= \ft(\M^{(k)}\y,\M^{(k)}\A).
\end{split}
\label{eq:ssdu_consistency}
\end{equation}

\vspace{0.5em}\noindent\textbf{Learning from multiple operators}\quad
These approaches leverage the fact that the image training set is seen through multiple $\A_i\sim\mathcal{A}$, where $\mathcal{A}$ is the operator set of same order as the mask set $\lvert\mathcal{M}\rvert={n\choose m}$. Multi-Operator Imaging \citep{tachella_unsupervised_2022} leverages this by drawing a random operator $\tilde\A\sim\mathcal{A}$ at each forward pass and regularises the MC loss with another loss that encourages consistency of $\ft$ over $\A_i$, shown to theoretically enable the function to learn in the null-space:

\begin{equation}
\L_\text{MOI}=\L_\text{MC}+\ell(\xhat,\ft(\tilde\A\xhat,\tilde\A)),\;\xhat=\ft(\y,\A)
    \label{eq:moi}.
\end{equation}

\citep{zhang_cycle-consistent_2024} constructs a very similar loss to MOI, enforcing multi-operator consistency on the measurement $\y$:
\begin{equation}
\L_\text{MOC-SSDU}=\L_\text{SSDU}+\ell(\y,\A\ft(\tilde\A\xhat,\tilde\A))
\label{eq:moc}.
\end{equation}

\vspace{0.5em}\noindent\textbf{Learning from invariance}\quad
Equivariant Imaging (EI) \citep{chen_equivariant_2021,chen_robust_2022,wang_perspective-equivariance_2025,wang_fully_2024} leverages the natural assumption that the image set $\mathcal{X}$ is invariant to a group $G$ of transformations $g\circ\x\in\mathcal{X},\forall\x\in\mathcal{X},g\in G$. The image set can then be interpreted as being observed through a set of multiple transformed operators $\A\circ g(\cdot)$ of order $\lvert G\rvert$, allowing the solver to ``see'' into the null-space. The assumption is constrained using:

\begin{equation}
\L_\text{EI}=\L_\text{MC}+\ell(\Tg\xhat,\ft(\A\Tg\xhat,\A)),\;\xhat=\ft(\y,\A)
\label{eq:ei},
\end{equation}

\noindent where $\Tg$ is an action of $G$, where, for MRI images, we use 2D rotations for $G$ following \citet{chen_robust_2022}. 

\vspace{0.5em}\noindent\textbf{Combining two methods}\quad
Our benchmark makes a new loss function apparent, which we can rapidly implement and test. Since adding more, well-targeted regularisation typically improves the stability and generalisation of the model, and that the operators leveraged by MOI (multiple physical operators) and EI (multiple virtual operators) are complementary, we propose a hybrid Multi-Operator Equivariant Imaging loss that unifies these strategies.
Assume that the image set is imaged by the \textit{augmented operator set} $\mathcal{A}_G=\{\A_i\Tg\;\forall \A_i\in\mathcal{A},g\in G\}$, the orbit space of $\mathcal{A}$ under $G$. Since $\lvert\mathcal{A}_G\rvert\approx\lvert\mathcal{A}\rvert\lvert G\rvert$ is much larger than in MOI or EI, we expect to provide more regularisation and thereby improve the performance. We leverage the augmented operator set using the loss function:

\begin{equation}
    \L_\text{MO-EI}=\L_\text{MC}+\ell(\Tg\xhat,\ft(\tilde\A_g\xhat,\tilde\A)),\;\tilde\A_g\sim\mathcal{A}_G
    \label{eq:moei}.
\end{equation}

For $G$, since we have soft deformable tissue \citep{gan_deformation-compensated_2022}, we expect $\mathcal{X}$ be invariant to the group of perturbative $C^1$-diffeomorphisms. See \cref{sec:loss_implementations} for details.

Another possible combination, was proposed in Equivariant Splitting \citep{sechaud_equivariant_2025} that circumvents the problem of measurement splitting methods when the splitting masks do not span the full domain in expectation, by constraining equivariance of $\ft$, enlargening the set of operators virtually:

\begin{equation}
    \L_\text{ES}=\mathbb{E}_g\L_\text{SSDU} \quad\text{where} \quad \A\rightarrow\A\Tg
    \label{eq:es}.
\end{equation}

A similar combination of the above was proposed in Equivariance2Inverse \citep{schut_equivariance2inverse_2025}, which simply takes a sum of the measurement splitting (\cref{eq:splitting}) and EI (\cref{eq:ei}) losses.

\vspace{0.5em}\noindent\textbf{Data augmentation}\quad
EI is related to data augmentation \citep{chen_imaging_2023}, which, when GT $\x$ is available, can be used to constrain invariance or equivariance on $\x$ \citep{fabian_data_2021}. Data augmentation consistency (DAC) methods \citep{xie_unsupervised_2020} such as VORTEX \citep{desai_vortex_2022} build on this in the semi-supervised context by performing data augmentation on $\y$ when GT is not available, chiefly to provide robustness to OOD data:

\begin{equation}
\L_\text{VORTEX}=
\begin{cases} 
\L_\text{sup}(\ft(\y,\A),\x) & \text{if GT exists} \\
\ell(\T_2\ft(y,\A),\ft(\T_1\y,\A\T_2^{-1})) & \text{otherwise}
\end{cases}
\label{eq:vortex}
\end{equation}

\noindent where $\T_1,\T_2$ are random transforms in $k$-space (\eg add noise, phase shift) and image space respectively. We test the ability of DAC methods in the fully GT free case, replacing $\L_\text{sup}$ with $\L_\text{MC}$ to enforce MC.

\vspace{0.5em}\noindent\textbf{Adversarial losses}\quad
Adversarial losses have been proposed to reconstruct MRI without GT via the dual training of $\ft$ and a discriminator $D$. While unconditional, generative adversarial networks \citep{bora_ambientgan_2018} are not feedforward at inference time, conditional methods using an adversarial loss \citep{cole_fast_2021}:

\begin{equation}
\L_\text{adversarial}=D(\tilde\A\ft(\y,\A),\tilde\y)
\label{eq:cole},
\end{equation}

\noindent where $\tilde\y\sim\mathcal{Y}$ \ie another randomly sampled ``real'' measurement from the training dataset $\mathcal{Y}$. The aim of $D$ is to distinguish between reconstructed measurements and training measurements, such that the aim of $\ft$ is to reconstruct high quality measurements that are indistinguishable from ``real'' measurements. Note that it is unclear how this loss can learn the unknown signal model and recover information from the null-space. UAIR \citep{pajot_unsupervised_2018} is a related method originally proposed for inpainting, using a multi-step feedforward loss consisting of adversarial and MC terms:

\begin{equation}
    \hat\y=\tilde\A \ft(\y,\A),\;\L_\text{UAIR}=D(\hat\y,\y)+\ell(\tilde\A\ft(\hat\y,\tilde\A),\hat\y)
    \label{eq:uair}.    
\end{equation}

\vspace{0.5em}\noindent\textbf{Joint reconstruction and denoising}\quad
For Scen. 2 (noisy), we consider composite losses that jointly learn to reconstruct and denoise from partial noisy measurements $\sigma>0$ alone (note that denoising can always independently be done as a separate preprocessing step \citep{aali_gsure_2024}). Several self-supervised denoising losses exist in the literature \citep{tachella_unsure_2025}; most combinations of these with reconstruction losses are yet to be explored. Stein's Unbiased Risk Estimator (SURE) \citep{stein_estimation_1981,ramani_monte-carlo_2008} can be used in place of $\L_\text{MC}$, shown to be an unbiased estimator of $\L_\text{MC}$ \citep{chen_robust_2022}. It can be seen as further ``recorrupting'' data:

\begin{equation}
\L_\text{SURE}=\L_\text{MC}+\frac{2\sigma^2}{\tau}\mathbf{b}^{\top}(\A\ft(\y+\tau\mathbf{b},\A) -A\ft(\y,\A))-\sigma^2
\label{eq:sure},
\end{equation}

\noindent where $\mathbf{b}\sim\mathcal{N}(\mathbf{0},\mathbf{I})$, $\tau$ is a hyperparameter and $\mathbb{E}_\y[\L_\text{SURE}]=\mathbb{E}_{\x,\y}[\ell(\A\ft(\y),\A\x)]$. Robust-EI \citep{chen_robust_2022} then simply use $\L_\text{Robust-EI}=\L_\text{SURE}+\L_\text{EI}$ to train. SURE has been generalised to the case of rank-deficient $\A$ in GSURE \citep{eldar_generalized_2009} and LDAMP-SURE \citep{metzler_unsupervised_2020,zhussip_training_2019}. However, ENSURE \citep{aggarwal_ensure_2023} is shown to improve on these by attempting to leverage multi-operator information of $\mathcal{A}$:

\begin{equation}
\begin{split}
\L_\text{ENSURE} &= \lVert \mathbf{R}(\A^\dagger\y - \ft(\y,\A))\rVert_2^2 \\[-2pt]
&+ \frac{2\sigma^2}{\tau}\mathbf{b}^{\top}
\!\left(\ft(\AT\y+\tau\mathbf{b},\A) - \ft(\AT\y)\right)
\end{split}
\label{eq:ensure}
\end{equation}

\noindent where $\mathbf{R}=\mathbb{E}\left[\mathbf{P}\right]^{-\frac{1}{2}}\mathbf{P},\mathbf{P}=\A^\dagger\A$. Note that when  $\sigma\rightarrow0$, ENSURE simply reduces to $\L_\text{MC}$. 

SSDU-style methods have also been adapted to the noisy case using Noisier2Noise-style recorruption losses \citep{moran_noisier2noise_2019}. Robust-SSDU \citep{millard_clean_2024} constructs a loss attempting to remove an additional recorruption $\tilde{\y}\sim\mathcal{N}(\y,\alpha^2\sigma^2\mathbf{I})$:

\begin{equation}
\begin{split}
\L_\text{Robust-SSDU} &= \L_\text{Weighted-SSDU}(\tilde{\y};\y) \\[-2pt]
&\quad + \lVert(1+\tfrac{1}{\alpha^2})\M^{(1)}\M(\A\ft(\tilde{\y},\A)-\y)\rVert_2^2
\end{split}
\label{eq:robust_ssdu}
\end{equation}

\noindent where $\alpha$ is a hyperparameter. Similarly, Noise2Recon-SSDU \citep{desai_noise2recon_2022} appends to SSDU a recorruption constructed as a noising special-case of VORTEX \citep{desai_vortex_2022} $\L_\text{Noise2Recon-SSDU}=\L_\text{SSDU}+\L_{\text{VORTEX}}$ where $\T_1=\mathcal{N},\T_2=\mathbf{I}$. DDSSL \citep{quan_dual-domain_2022} also adds a random noise special-case of $\mathcal{L}_{\text{EI},\mathbf{T}=\mathcal{N}}$ to the Recorrupted2Recorrupted loss \citep{pang_recorrupted--recorrupted_2021}:

\begin{equation}
\L_\text{R2R}=\lVert\A\ft(\y+\alpha\omega,\A)-\left(\y-\frac{\omega}{\alpha}\right)\rVert_2^2
\label{eq:r2r}.
\end{equation}

\vspace{0.5em}\noindent\textbf{Dynamic losses}\quad
For Scen. 6 (dynamic MRI), we adapt the losses to learn from temporal patterns. For SSDU losses we let $\M_1$ vary randomly in time \citep{acar_self-supervised_2021}, and DDEI \citep{wang_fully_2024} adds temporal symmetries to the EI group $\mathcal{G}$.

\subsection{Related work}

\noindent\textbf{Surveys for imaging inverse problems}\quad Recent surveys exist covering inverse problems with deep learning \citep{ongie_deep_2020}, for MRI reconstruction \citep{heckel_deep_2024,chen_ai-based_2022} and without GT \citep{zeng_review_2021,akcakaya_unsupervised_2022}. While \citep{chen_ai-based_2022,heckel_deep_2024,ongie_deep_2020,akcakaya_unsupervised_2022,xu_towards_2026} provide valuable general overviews, they discuss only a subset of existing self-supervised methods (\eg zero to five), or methods from only one category mentioned in \cref{sec:benchmark}, such as measurement splitting. While GT-free MRI surveys \citep{zeng_review_2021,akcakaya_unsupervised_2022} highlight the merits of self-supervised imaging, they lack a principled comparison of the core theoretical differences between methods, comparing instead entire pipelines, conflating therefore mostly orthogonal problems including model architecture $\ft$ or forward operator $\A$. Furthermore, no experimental results are provided by \citep{heckel_deep_2024,zeng_review_2021,akcakaya_unsupervised_2022}. Previous individual methodological works either do not report comparison with other methods \eg \citep{cole_fast_2021,chen_robust_2022}, or compare to very few \citep{millard_clean_2024,desai_vortex_2022,aggarwal_ensure_2023}, limiting their applicability; a comprehensive benchmark comparison is lacking. Papers commonly do not provide public implementation details of competitor methods, potentially reducing the trustworthiness of the comparisons, especially when competitors' codebases are directly used without disentangling the method from the experimental setup $(\ft,\A)$. On the contrary, we highlight the key conceptual difference, the loss function, and benchmark this while keeping constant all other elements of the imaging pipeline, providing benchmark results across multiple fixed setup scenarios $(\ft,\A)$. Furthermore, we reimplement the compared methods clearly in a common codebase, bypassing the practical barriers of code uninteroperability, to encourage reproducible research.

\vspace{0.5em}\noindent\textbf{Benchmarks for imaging inverse problems}\quad Numerous imaging benchmarks on real data have been proposed for tasks such as MRI or CT \citep{zbontar_fastmri_2018,desai_skm-tea_2022,kiss_benchmarking_2025,zheng_inversebench_2025}. However, to the best of our knowledge, in any domain, no benchmarks on image reconstruction from partial measurements without GT exist. Such a benchmark is crucial for enabling fair, reproducible comparison of self-supervised reconstruction methods and driving progress toward reliable GT-free imaging.

%% file: sec/3_experiments.tex
\section{Experimental setup}
\label{sec:experiments}

\input{sec/results_table1}

\input{sec/results_table2}

We evaluate all \cref{tab:methods} methods using our reimplementations across the various scenarios. Each scenario uses a fixed setup across all methods, and this setup varies across scenarios, including varied models, anatomy, dataset, scanner vendor, acceleration and acquisition noise.

\vspace{0.5em}\noindent\textbf{Data}\quad
For scenarios 1-4 we retrospectively simulate $\y$ from slices of the popular FastMRI brain dataset \citep{zbontar_fastmri_2018}, in order to fully control the forward operator $\A$ (its distribution is needed for \citep{yaman_multi-mask_2022,millard_theoretical_2023,hendriksen_noise2inverse_2020,tachella_unsupervised_2022}) and have GT $\x$ for evaluation. The specific forward operator $\A$ varies for each benchmarked scenario. For the multi-coil case we assume known coil maps to focus on the reconstruction task and not introduce artifacts associated with map estimation. For external data Scenarios 5, 6 \& 7, we use respectively: raw multicoil sagittal knee $k$-space slices from the SKM-TEA \citep{desai_skm-tea_2022} test set with estimated coil maps via JSENSE \citep{ying_joint_2007}, raw prospectively-undersampled multicoil axial knee $k$-space slices from \citet{yu_validation_2022}, and cardiac 2D+t sequences from the CMRxRecon \citep{wang_cmrxrecon_2023} test set.

\vspace{0.5em}\noindent\textbf{Model $\ft$}\quad
From the vast architectural literature \citep{heckel_deep_2024} for scenarios 1-4 we choose an unrolled network architecture, which balances being specific for efficient image reconstruction yet sufficiently general to not favour specific losses. We choose the popular MoDL \citep{aggarwal_modl_2019}, unrolling finite (three) steps of half-quadratic splitting \citep{zhang_plug-and-play_2022}, with a U-Net backbone \citep{ronneberger_u-net_2015}. Identical hyperparameters are used for all loss functions to provide valid comparisons. For Scenarios 5 \& 7 we instead fine-tune a physics-informed foundation model architecture \citep{terris_reconstruct_2025} and for Scenario 6 we train a convolutional recurrent network \citep{qin_convolutional_2019}.

\vspace{0.5em}\noindent\textbf{Evaluation}\quad
We report PSNR and SSIM as used in FastMRI \citep{zbontar_fastmri_2018}, as well as LPIPS for perceptual quality \citep{zhang_unreasonable_2018}. We compare to the baseline $\A^\dagger\y$, corresponding to the zero-filled (ZF) in single-coil MRI and SENSE \citep{pruessmann_sense_1999} in multi-coil MRI (we avoid comparisons to iterative methods \citep{lustig_compressed_2008}). We also train a gold-standard model with the supervised oracle loss $\ell(\xhat,\x)$. Further setup details and loss and training code are in \cref{sec:experiment_details}. Benchmark code and contribution instructions are at \href{https://github.com/Andrewwango/ssibench}{github.com/Andrewwango/ssibench}.

\vspace{0.5em}\noindent\textbf{Extending}\quad
We highlight that our modular benchmark framework facilitates replacing any element of the setup used here: other SotA architectures, popular image quality metrics, or forward operator scenarios. To evidence ease of benchmark extensibility, further experiments investigate varying each benchmark module: pilot results on SotA GT-free pretrained diffusion models, testing performance scaling as unrolling depth increases or on different architectures (VarNet \citep{hammernik_learning_2018}, a flat CNN \citep{zhang_beyond_2017}, and transformer \citep{zamir_restormer_2022}), component ablations on MO-EI and SSDU, and ease of transferability to a sparse-view CT task and an environmental imaging task. 

\subsection{Results}
\label{sec:results}

\begin{figure*}[tb]
  \centering
  \includegraphics[width=0.999\textwidth]{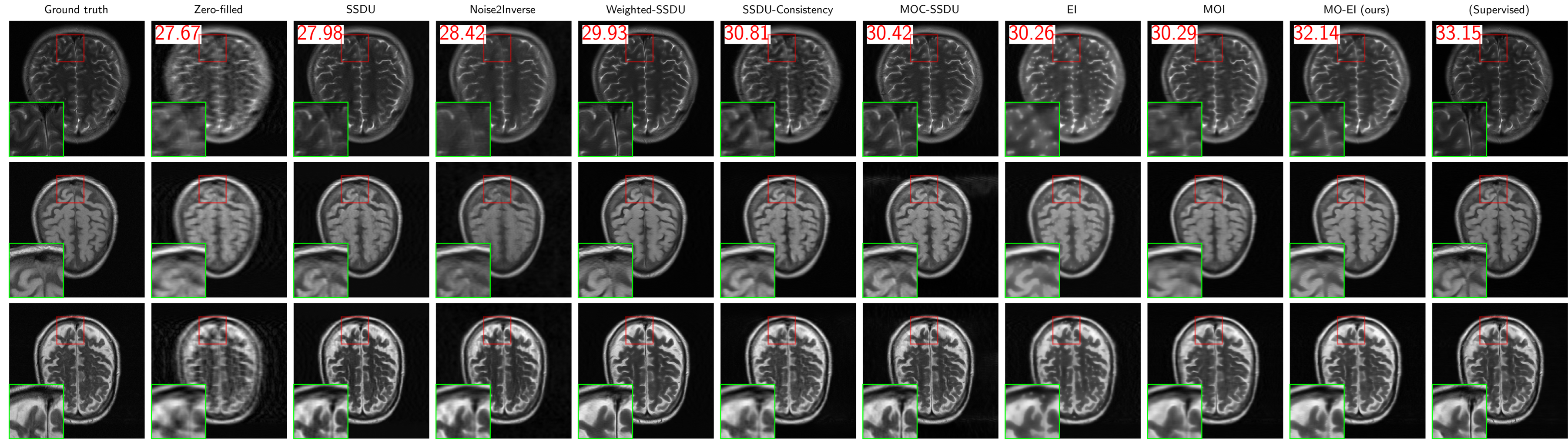}
  \caption{$6\times$ acc. (scenario 1) recons with test-set PSNR, showing selected highest-performing methods. See \cref{tab:results_scenario1345} for more.}
  \label{fig:results_scenario1}
\end{figure*}

\noindent\textbf{Scenario 1 (single-coil)}\quad
Results are shown in \cref{tab:results_scenario1345,fig:results_scenario1}. As expected, MC cannot improve the ZF performance, as \cref{eq:mc} cannot learn information from the null-space; results presented in \citep{senouf_self-supervised_2019} may instead heavily rely on a specific network's inductive bias. EI and MOI both show good results but suffer from strong artifacts in the brain images, but MOI slightly outperforms EI because its set of virtual operators in \cref{eq:moi} is likely larger than that of \cref{eq:ei}, so that the per-sample null-space is better covered per training step. MO-EI improves significantly on both these previous SotA methods (see \cref{sec:statsig} for statistical test), approaching the oracle supervised learning performance, as \cref{eq:moei} is able to recover information from the null-space via a larger set of virtual operators than that of EI or MOI. SSDU suppresses artifacts and recovers sharper edges quite well, but its quantitative performance is relatively poor, because \cref{eq:splitting} constructs an estimator conditioned on a small subset of the available measurements. Even though the loss is equal to the supervised loss associated with this masked subset in expectation, it is inefficient and thus may perform worse in finite data settings, and the estimator $\ft(\y,\A)$ over-estimates the MMSE estimator at inference-time. Other methods based on splitting improve on this, as they recover the MMSE at inference-time. We attribute the very poor performance of the adversarial losses (marked with $^*$) to the well-known instability and sensitivity of training $\ft$ and $D$ \citep{salimans_improved_2016}: after extensive experimentation we were unable to reproduce the results from \citep{cole_fast_2021} (which use a proprietary dataset), which may require heavy hyperparameter tuning and specific architectures. Since UAIR \citep{pajot_unsupervised_2018} does not show results on accelerated MRI, we assume their results do not transfer, and further work is needed to find better theoretically-justified adversarial implementation choices to reach stronger conclusions for adversarial methods. VORTEX learns only marginal information compared to MC, showing that measurement-domain data augmentations alone cannot recover information from the null-space.

\vspace{0.5em}\noindent\textbf{Scenario 2 (noisy)}\quad
Results are shown in \cref{tab:results_scenario2} and figure in \cref{fig:results_scenario2}. ENSURE denoises a moderate amount, but strong undersampling artifacts remain. Noise2Recon-SSDU and Robust-SSDU perform fairly well, resulting in clearer reconstructions with fairly sharp edges. The SURE Robust-EI/MO-EI methods are almost artifact-free with high metrics, because of the combination of null-space loss and SURE \cref{eq:sure} which approximates the clean supervised loss in expectation, but edges are less sharp due to the averaging. All loss functions without the denoising terms perform poorly as expected.

\begin{figure*}[tb]
    \centering
    \includegraphics[width=0.999\textwidth]{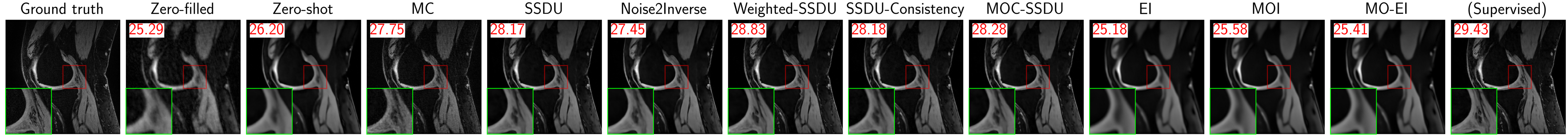}
    \caption{$32\times$ acc. SKM-TEA \citep{desai_skm-tea_2022} fine-tuning (scenario 5) recons, showing selected highest-performing methods. See \cref{tab:results_scenario1345} for more.}
    \label{fig:results_scenario5}
  \end{figure*}
\begin{figure*}[tb]
    \centering
    \includegraphics[width=0.999\textwidth]{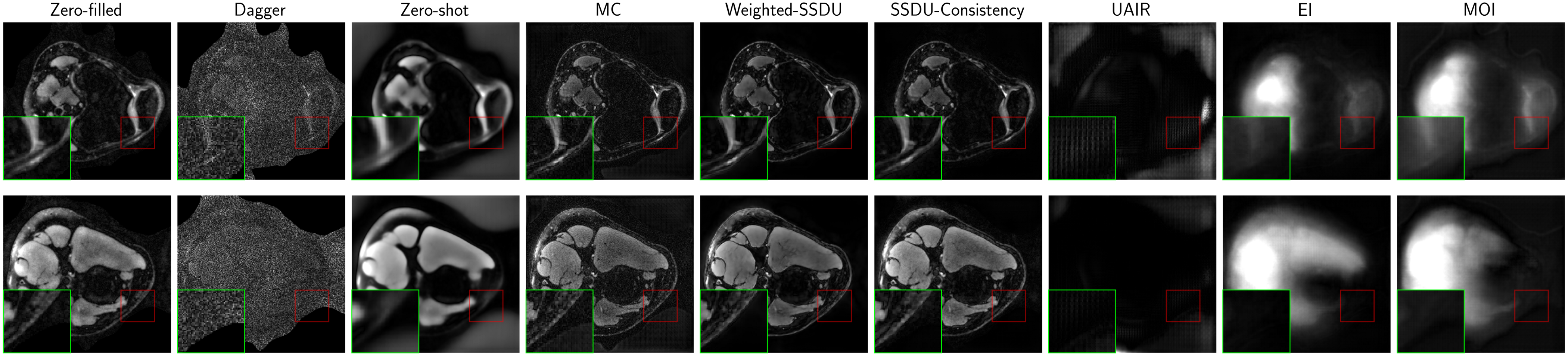}
    \caption{$7\times$ acc. prospectively undersampled knees \citep{yu_validation_2022} (scenario 7) : fine-tuning recons, showing selected highest-performing methods.}
    \label{fig:results_scenario7}
  \end{figure*}
  
\vspace{0.5em}\noindent\textbf{Scenario 3 (single-mask)}\quad
Results are shown in \cref{tab:results_scenario1345} and figure in \cref{fig:results_scenario3}. As expected, the best-performing method is EI, which recovers information from the null-space but is the only loss \textit{without} the assumption that the image set is seen through multiple masks. Other methods which do require this assumption still learn some information over the ZF solution, demonstrating the challenge of disentangling actual learning from network inductive bias.

\vspace{0.5em}\noindent\textbf{Scenario 4 (multi-coil)}\quad
Results are shown in \cref{tab:results_scenario1345} and figure in \cref{fig:results_scenario4}. As expected, MC learns the SENSE $\A^\dagger\y$, as it constructs an estimator that uses only the data fidelity term that is used in the pseudo-inverse. Interestingly, splitting-based methods consistently outpeform EI, MOI and MO-EI, with weighted-SSDU approaching supervised learning. While in previous scenarios, splitting methods struggled to recover information from larger null-spaces, here $\A$ has a higher effective rank (analysis in \cref{sec:svd}), which highlights instead their strength on sharp edges.

\vspace{0.5em}\noindent\textbf{Scenario 5 (fine-tuning)}\quad
Results are shown in \cref{tab:results_scenario1345,fig:results_scenario5}. The zero-shot base foundation model \citep{terris_reconstruct_2025}, trained with supervised learning, produces out-of-domain, cartoonish artifacts in the knees (\cref{fig:results_scenario5}, 3\textsuperscript{rd} image). Weighted-SSDU recovers much of the original sharp detail, as its loss \cref{eq:millard} is equal in expectation to the supervised loss. However, EI/MOI null-space losses now \textit{do not} improve the performance, since they cannot recover any more information from the null-space than that recovered by the well-trained MMSE foundation model. Interestingly, the foundation model's inductive bias allows pure MC to improve performance, although reconstructions remain noisy.

\vspace{0.5em}\noindent\textbf{Scenario 6 (dynamic)}\quad
Results are shown in \cref{tab:results_scenario6} and figure in \cref{fig:results_scenario6}. As before, EI/MOI-style losses recover information from the null-space and remove 10$\times$ undersampling artifacts. The addition of the diffeo transform for EI improves the results since the cardiac sequence displays strong deformable frame self-similarity.

\vspace{0.5em}\noindent\textbf{Scenario 7 (prospective)}\quad
Axial reconstructions are shown in \cref{fig:results_scenario7}; since no GT is present, we do not report quantitative results. We observe that the best methods recover sharp details in the knees, such as in the patellar cartilage or the muscle fascia, and remove the residual noise compared to the zero-filled reconstructions; since the noise level is relatively high, the SENSE reconstruction performs poorly. Similar to Scenario 5, only Weighted-SSDU recovers the fine cartilage detail, whereas the EI/MOI methods catastrophically fail in fine-tuning.

\vspace{0.5em}\noindent\textbf{Further results}\quad
Further experiments and ablations are reported in \cref{sec:further_experiments}.
\textbf{Ablation}: best performing implementation choices correspond to the theoretically optimal choices, while performance, and thus our conclusions, are sensitive to suboptimal choices.
\textbf{Diffusion} (\cref{tab:results_diffusion}): posterior sampling methods achieve good perceptual but poorer distortion performance compared to MMSE estimators as expected (these models were \textit{pretrained on much larger datasets} so results should not be directly compared). \textbf{Unrolling}: performances scale uniformly with unrolling depth, suggesting that our results would scale to setups with deeper models.
\textbf{Model architecture}: with either a 50$\times$ smaller architecture and flat CNN, or shallower transformer-based architecture, Scenario 1 performance decreases but ranking is broadly preserved, suggesting that loss choice is orthogonal to architecture design for this scenario.

%% file: sec/results_table1.tex
\begin{table*}[tb]
\centering
\caption{Benchmarking test set results ($\mu\pm\sigma$) for scenarios 1, 3, 4 and 5. \textbf{Best unsupervised method in bold}.}
\label{tab:results_scenario1345}
\resizebox{\textwidth}{!}{
\begin{tabular}{lccc|cc|cc|ccc}
\hline
\multicolumn{1}{r}{{Scenario}} &
\multicolumn{3}{c}{\textbf{(1)} Brain MRI 6$\times$ acc.} &
\multicolumn{2}{c}{\textbf{(3)} Single-mask} &
\multicolumn{2}{c}{\textbf{(4)} Multicoil $C=4$} &
\multicolumn{3}{c}{\textbf{(5)} SKM-TEA  knee fine-tune 32$\times$} \\
\cline{2-11}
Loss & PSNR$\uparrow$ & SSIM$\uparrow$ & LPIPS$\downarrow$

 & PSNR$\uparrow$ & SSIM$\uparrow$
 & PSNR$\uparrow$ & SSIM$\uparrow$
 & PSNR$\uparrow$ & SSIM$\uparrow$ & LPIPS$\downarrow$ \\ \hline

Zero-shot & -- & -- & -- & -- & -- & -- & -- & 26.20\std{.54} & .5081\std{.0226} & .5273\std{.0379} \\
Zero-filled & 27.67\std{2.40} & .7862\std{.05} & .3270\std{.0426} & 28.02\std{2.26} & .7900\std{.05} & 27.82\std{2.41} & .7988\std{.05} & 25.29\std{.50} & .5896\std{.02} & .4913\std{.0257} \\
MC & 27.66\std{2.40} & .7858\std{.0550} & .3273\std{.0426} & 28.02\std{2.26} & .7900\std{.0496} & 28.96\std{2.59} & .8271\std{.0495} & 27.75\std{.65} & \textbf{.5965}\std{.0277} & \textbf{.2713}\std{.0235} \\
SSDU & 27.98\std{1.43} & .7485\std{.0667} & .1965\std{.0417} & 21.89\std{1.07} & .6288\std{.0510} & 31.47\std{2.67} & .8705\std{.0534} & 28.17\std{.62} & .5800\std{.0174} & .4265\std{.0371} \\
Noise2Inverse & 28.42\std{2.02} & .7853\std{.0735} & .2347\std{.0416} & 24.63\std{2.07} & .6559\std{.0676} & 30.93\std{2.65} & .8589\std{.0570} & 27.45\std{.64} & .5465\std{.0175} & .4509\std{.0367} \\
Weighted-SSDU & 29.93\std{1.66} & .8355\std{.0626} & \textbf{.1304}\std{.0434} & 30.14\std{1.64} & .8454\std{.0615} & \textbf{33.03}\std{2.36} & .8991\std{.0479} & \textbf{28.83}\std{.60} & .5901\std{.0180} & .3964\std{.0394} \\
SSDU-Consistency & 30.81\std{2.58} & .8495\std{.0581} & .1845\std{.0370} & 31.05\std{2.50} & .8614\std{.0567} & 32.30\std{2.58} & .8949\std{.0425} & 28.18\std{.61} & .5900\std{.0167} & .4088\std{.0304} \\
MOC-SSDU & 30.43\std{2.71} & .8198\std{.0854} & .1578\std{.0568} & 27.85\std{1.86} & .7717\std{.0833} & 31.80\std{2.69} & .8761\std{.0529} & 28.28\std{.61} & .5823\std{.0170} & .4238\std{.0375} \\
Adversarial & 18.52\sstd{.31} & .4732\std{.0388} & .3927\std{.0413} & 26.53\std{1.62} & .7013\std{.0370} & 17.47\sstd{1.93} & .6464\std{.0590} & 15.60\sstd{1.06} & .3355\std{.0268} & .6608\std{.0410} \\
UAIR & 14.00\sstd{1.88} & .3715\std{.0824} & .4826\std{.0584} & 18.44\sstd{1.61} & .5388\std{.0542} & 15.26\sstd{2.16} & .3453\std{.0540} & 16.07\sstd{.40} & .3902\std{.0231} & .7461\std{.0778} \\
VORTEX & 27.75\std{2.35} & .7898\std{.0543} & .3201\std{.0405} & 28.07\std{2.26} & .7916\std{.0507} & 23.59\std{1.17} & .5846\std{.0469} & 25.01\std{1.07} & .5766\std{.0291} & .5517\std{.0508} \\
ES & 29.80\std{2.09} & .8105\std{.0591} & .3309\std{.0508} & 30.26\std{1.97} & .8242\std{.0644} & 30.82\std{2.48} & .8370\std{.0653} & 25.96\std{.90} & .5150\std{.0179} & .4354\std{.0308} \\
E2I & 29.92\std{1.89} & .8122\std{.0587} & .2972\std{.0516} & 25.45\std{1.10} & .6760\std{.0435} & 32.12\std{2.67} & \textbf{.9116}\std{.0590} & 27.68\std{.61} & .5500\std{.0176} & .4696\std{.0329} \\
EI & 30.26\std{2.61} & .8523\std{.0542} & .2429\std{.0423} & \textbf{31.99}\std{2.80} & \textbf{.8806}\std{.0486} & 31.66\std{2.74} & .8769\std{.0494} & 25.18\std{.56} & .4676\std{.0194} & .5952\std{.0342} \\
MOI & 30.29\std{2.88} & .8651\std{.0528} & .2341\std{.0472} & 31.60\std{2.74} & .8789\std{.0477} & 31.37\std{2.86} & .8810\std{.0502} & 25.58\std{.57} & .4801\std{.0199} & .5742\std{.0354} \\
MO-EI & \textbf{32.14\std{2.73}} & \textbf{.8846}\std{.0498} & .1803\std{.0406} & 31.11\std{2.69} & .8713\std{.0496} & 31.56\std{2.85} & .8836\std{.0477} & 25.41\std{.57} & .4756\std{.0200} & .5841\std{.0374} \\
(Supervised) & 33.15\std{2.76} & .9032\std{.0435} & .1185\std{.0335} & 34.03\std{2.49} & .9040\std{.0435} & 33.89\std{2.78} & .9147\std{.0365} & 29.43\std{.59} & .6203\std{.0142} & .3259\std{.0382} \\ \hline
\end{tabular}
}
\end{table*}

%% file: sec/results_table2.tex
\begin{table*}[h]
    \centering
    \begin{minipage}[t]{0.48\textwidth}
    \centering
      \caption{\textbf{Scenario 2} (joint denoising \& recon) test results.}
      \label{tab:results_scenario2}
    \resizebox{0.96\textwidth}{!}{
    \begin{tabular}{llll}
    \hline
    Loss & PSNR $\uparrow$ & SSIM $\uparrow$ & LPIPS$\downarrow$ \\ \hline
Zero-filled & 24.34\std{1.01} & .4428\std{.03} & .4623\std{.0530} \\
ENSURE & 26.29\std{1.51} & .5856\std{.0360} & .3117\std{.0474} \\
Robust-SSDU & 27.42\std{1.13} & .6159\std{.0425} & \textbf{.3040}\std{.0589} \\
Noise2Recon-SSDU & 27.84\std{1.78} & .7661\std{.0727} & .3267\std{.0601} \\
DDSSL & 28.25\std{2.30} & .7836\std{.0548} & .3786\std{.0652} \\
ES-R2R & 28.14\std{1.51} & .6377\std{.0381} & .3782\std{.0462} \\
E2I & 25.80\std{.80} & .4668\std{.0471} & .4681\std{.0540} \\
Robust-EI & 29.07\std{2.37} & .8227\std{.0578} & .3405\std{.0621} \\
Robust-MO-EI & \textbf{29.72}\std{2.44} & \textbf{.8409}\std{.0575} & .3311\std{.0614} \\
Non-robust MO-EI & 26.12\std{1.92} & .6002\std{.0602} & .3865\std{.0518} \\
Non-robust Weighted-SSDU & 25.91\std{.94} & .5477\std{.0511} & .3174\std{.0734} \\
(Supervised) & 30.19\std{2.10} & .8411\std{.0552} & .2532\std{.0562} \\
    \hline
      \end{tabular}
      }
    \end{minipage}%
    \hspace{0.01\textwidth}
    \begin{minipage}[t]{0.48\textwidth}
      \centering
      \caption{\textbf{Scenario 6} ($10\times$ acc. dynamic cardiac MRI).}
      \label{tab:results_scenario6}
      \resizebox{0.8\textwidth}{!}{
      \begin{tabular}{lllll}
      \hline
      Loss & PSNR $\uparrow$ & SSIM $\uparrow$ & LPIPS$\downarrow$ \\ \hline
Zero-filled & 30.41\std{1.07} & .8041\std{.03} & .2137\std{.0248} \\
MC & 30.41\std{1.07} & .8041\std{.0299} & .2137\std{.0248} \\
t-SSDU & 30.18\std{1.38} & .8212\std{.0272} & .1992\std{.0272} \\
Weighted-t-SSDU & 29.65\std{1.56} & .7836\std{.0290} & .1901\std{.0274} \\
t-SSDU-Consistency & 31.82\std{1.15} & .8477\std{.0236} & .1714\std{.0225} \\
MOC-t-SSDU & \textbf{31.88}\std{1.19} & .8485\std{.0235} & \textbf{.1572}\std{.0192} \\
Diffeo-EI & 31.81\std{1.39} & \textbf{.8548}\std{.0244} & .1691\std{.0247} \\
EI & 31.22\std{1.20} & .8431\std{.0247} & .1931\std{.0246} \\
MOI & 31.47\std{1.25} & .8471\std{.0239} & .1938\std{.0259} \\
DDEI & 31.36\std{1.17} & .8450\std{.0248} & .1853\std{.0245} \\
(Supervised) & 33.95\std{1.75} & .8750\std{.0244} & .1038\std{.0227} \\
      \hline
      \end{tabular}
      }
    \end{minipage}
    \end{table*}

%% file: sec/4_discussion.tex
\section{Discussion and Conclusions}


\vspace{0.5em}\noindent\textbf{Design} We close an important gap in self-supervised imaging (SSI) for MRI reconstruction without GT by designing a modular and comprehensive benchmark, \textbf{SSIBench}, that unifies 21 distinct end-to-end training losses.
This systematic comparison is crucial for targeted future research and industrial adoption. We formulate seven realistic MRI scenarios that each test different capabilities of methods in diverse acquisition settings, architectures and anatomies.

\vspace{0.5em}\noindent\textbf{Insights} Our extensive experiments on standardised setups show a wide range in performance, with different methods leading (\eg by ~2 dB) on different scenarios and metrics. A possible decision framework for practitioners, based on the results of the tested scenarios, is:

\begin{itemize}
    \item Highly ill-posed ($\alpha\gg C$): use MO-EI, MOI or EI;
    \item Multicoil ($\alpha\lesssim C$): use weighted-SSDU;
    \item Fine-tuning pretrained models: use weighted-SSDU;
    \item Sampling pattern is fixed for all patients: use EI;
    \item High acquisition noise: use above losses with appended denoising terms such as SURE.
\end{itemize}

These observations motivate further ML research by exposing the ongoing need for a reliable method that can lead on all scenarios. The benchmark facilitates the systematic combination and prototyping of loss components to discover more effective configurations such as MO-EI, paving the way for future exploration of even more promising ones.

Although one might suspect that fixing the model hides architectural bias, further experiments (\cref{sec:further_experiments}) suggests that Scenario 1 performance trends do generalise across model architectures and sizes and their choice is thus orthogonal to the loss function for this scenario.

\vspace{0.5em}\noindent\textbf{Limitations}
In future we aim to include a wider range of datasets \eg raw noisy low-field data \citep{lyu_m4raw_2023}, other forward operators $\A$ \eg non-Cartesian sampling and more complex models $\ft$ \citep{tanabene_benchmarking_2024}, a larger set of \citep{desai_skm-tea_2022,breger_study_2025}, or other realistic MRI subtasks \eg joint coil map estimation \citep{hu_spicer_2024}. Toward further expansion on these fronts, our modular setup facilitates swapping out any of these components to accommodate diverse research configurations. We demonstrate this capability by extending the benchmark framework beyond MRI and provide a proof-of-concept example benchmarking SSI for a task in environmental imaging and CT in \cref{sec:hsi}.

Furthermore, while end-to-end trained methods do not require large datasets, long inference times nor per-image training, and thus favour clinical adoption, SSIBench is limited by not including other promising methods that today are not yet clinically feasible, but may soon become so. Future work thus would add these nascent methods, such as few-step diffusion models or generalisable implicit neural representations. 

We note that research in self-supervised imaging is an active and fast-moving field. Therefore, the paper inevitably presents a snapshot of methods taken at the time of writing. Crucially, the creation of the modular open-source benchmark allows future methods to be added and extended as new methods become practical.

\vspace{0.5em}\noindent\textbf{Benchmark resource and outlook} Our benchmark site allows researchers to easily use reimplementations of all losses and contribute. This lowers the barrier to entry and enables ML researchers and practitioners to a) contribute and compare new ML methods on a standard evaluation framework, and b) rapidly and fairly test these methods on their own setups. We encourage the community to generalise our insights and use our benchmark resource as a blueprint to unlock self-supervised learning for emerging, exciting domains within MRI and beyond, while remaining mindful of potential misuse of high quality images.

%% file: sec/X_suppl.tex
\clearpage
\setcounter{page}{1}
\appendix

\section{Further experimental details}
\label{sec:experiment_details}
\subsection{Data preprocessing details}
We use a subset of the FastMRI \citep{zbontar_fastmri_2018} dataset for our experiments, and our dataset is reproducible by following the instructions given on the benchmark website. We take the middle axial slice of 455 brain volumes, use the provided $320\times320$ root-sum-square reconstructions as GT, normalise them to $[0,1]$ and retrospectively simulate measurements using an undersampled Fourier transform using the DeepInverse library \citep{tachella_deepinverse_2025}. We randomly create train-test datasets with 80-20 split, since the original FastMRI test set does not have GT for evaluation. For the subsampling masks $\M_i\sim\mathcal{M}$, we use random 1D Gaussian masking with $6\times$ acceleration and 8\% autocalibration (ACS) lines kept in the centre. For the noisy scenario we add i.i.d.\ Gaussian measurement noise with $\sigma=0.1$, and for the single-operator scenario we fix one $\M\sim\mathcal{M}$. For the multi-coil scenario we assume known sensitivity coil maps to decouple the parameter estimation problem from reconstruction, and assume a fixed number of coils $C$ so that the number of measurements $m$ is constant over the dataset. This is necessary as coil sensitivity estimation introduces its own artifacts, particularly in low image-density areas outside the main region of interest, which we do not wish to introduce into our setup. Therefore, we simulate coil maps using \verb+sigpy+ \citep{ong_sigpy_2019} with $C=4$ and simulate measurements using \cref{eq:mri_problem}.

The SKM-TEA \citep{desai_skm-tea_2022} dataset contains raw quantitative (qDESS) acquisitions from a GE MR750 3T scanner (GE, Waukesha, WI) and estimated coil maps via JSENSE \citep{ying_joint_2007}. We take raw multicoil data from 100 slices across 5 volumes and retrospectively simulate 2D Poisson disk undersampling at 32$\times$ acc. as per the original paper. The CMRxRecon \citep{wang_cmrxrecon_2023} dataset contains cardiac cine sequences of length 12. We take raw data from 100 slice sequences from the training set, split by patient ID, and apply k-t-space Gaussian undersampling at 10$\times$ acc. For dynamic MRI we swap the model $\ft$ for the CRNN \citep{qin_convolutional_2019}. The knee volume in \citet{yu_validation_2022} is prospectively acquired from a Siemens MAGNETOM Prisma Fit 3T scanner (Siemens Healthcare, Erlangen, Germany) with a T2 SPACE sequence using a 7$\times$ 3D Poison disk mask with 18 coils; the coil maps are estimated with ESPIRiT \citep{uecker_espiriteigenvalue_2014}. We take slices in the axial dimension for the 2D experiment.

Note that while retrospective simulation is necessary for a controlled benchmarking experiment, future experiments would evaluate benchmarked methods on prospectively acquired data \citep{shimron_implicit_2022,chaudhari_prospective_2021,yu_validation_2022}. 

Note that all images are normalised to the range $[0,1]$ before plotting.

\subsection{Model and hyperparameter details}

The unrolled network $\ft$ unrolls the iterative bi-level optimisation problem provided by half-quadratic splitting \citep{zhang_plug-and-play_2022} for $u$ iterations, originally proposed for MRI in \citep{aggarwal_modl_2019}, and we take $u=3$. During inference, this model $\ft$ requires $u$ forward passes of the denoiser. For the denoiser of our unrolled network, we use a residual U-Net \citep{ronneberger_u-net_2015} with no batch norm of depth 4, with a total of 8.6M parameters. We train with the Adam optimiser at $1e-3$ learning rate, and we step this per method down to $1e-5$ to achieve convergence. All models are trained with batch size of 4 using 1 NVIDIA GeForce RTX 3090, taking around 15GB memory, with an average training time of 6 hours. Average training time per batch is 0.5s, and average inference time of the trained models $\ft$ is 0.1s per batch.

Experimental results (mean, sample standard deviation) are reported on held-out test sets for each dataset. Since end-to-end trained models are deterministic (\ie their input is the measurement, rather than a random latent), inference runs are not repeated. Simulation of retrospective random undersampling masks, random noise, dataset random split and model initialisation used fixed seeds for all experiments to keep the setup fixed and repeatable across runs. Note that the reported results may be a function of the seeds used during training and are thus indicative. Furthermore, fixing certain hyperparameters in the architecture, loss functions and training recipe potentially affects the rankings of some methods across scenarios; future work would perform more thorough hyperparameter search across these.

\subsection{Individual loss implementation details}
\label{sec:loss_implementations}

We provide details of the benchmarked loss functions' reimplementations below. All components of our modular experiments are implemented using DeepInverse \citep{tachella_deepinverse_2025}, and full code is at \href{https://github.com/Andrewwango/ssibench}{github.com/Andrewwango/ssibench}. 

\paragraph{General notes} Some losses compute additional forward passes of $\ft$ per training iteration. For these, more Monte Carlo passes per iteration could improve performance but we uniformly limit this to one additional pass for all methods for efficiency. Secondly, the training metric $\ell$ in benchmarked methods' original papers varies between L1 and L2 (MSE); since this is not instrumental in recovering information, and many methods' theory rely on the MSE, we use L2 for standardised evaluation. Thirdly, where there is more than one term in the loss we weight the terms evenly.

\paragraph{Measurement consistency:} We simply take $\ell$ as the MSE in kspace as per \citep{senouf_self-supervised_2019} with $\beta>0,\alpha=\gamma=0$. In the multi-coil case, we take $\ell$ as the MSE in the backprojected $\A^\top$ space as we find this provides better results.

\paragraph{Measurement splitting:} We use a 2D mask with split ratio $\rho=0.6$ following \citet{yaman_self-supervised_2020}, drawn randomly each instance during training following \citet{yaman_multi-mask_2022}. For weighted-SSDU we use an independent 1D Gaussian mask following \citet{millard_theoretical_2023}. At inference time, we do not perform additional correction for both \citet{millard_theoretical_2023,millard_clean_2024} since our network maps directly to images $\ft:\mathcal{Y}\rightarrow\mathcal{X}$. We ablate and justify the splitting mask choices in \cref{sec:ssdu_ablation}. For \citep{hu_self-supervised_2021}, the authors train two networks, but we follow their ablation and share these weights to keep the number of parameters constant for a fair setup, follow the partitioning used above, and remove the ISTA-Net-specific loss terms. Note that \citep{wang_parcel_2023} is equivalent to the above but takes the consistency loss on embeddings of $\xhat_1,\xhat_2$ instead. Both \citep{wang_parcel_2023,zhou_dual-domain_2022} are equivalent to the above but with the addition of a superfluous $\mathcal{L}_\text{MC}$.

\begin{proposition}[Equivalence of unpaired Artifact2Artifact as sum of SSDU losses]
Consider the Artifact2Artifact \citep{liu_rare_2020} scenario where $\y_i^{(k)}$ are independent sets of measurements of the same subject $i$ (\eg arriving in a stream) such that $\y_i=\bigcup_{k}\y_i^{(k)}\in\mathbb{R}^m$ is the full undersampled measurement from one acquisition $i$ (i.e. there are no more measurements of the same subject). $\A_i^{(k)\top}\y_i^{(k)}$ are then the independent ``artifact''-corrupted images. The Artifact2Artifact loss draws a pair $k,l$ randomly at each iteration and constructs the loss
\begin{equation*}
    \L_\text{A2A}^{(k,l)}=\ell(\A_i^{(l)}\ft(\y_i^{(k)},\A_i^{(k)}),\y_i^{(l)}),
\end{equation*}
where $\A_i^{(k)},\A_i^{(l)}$ are the operators associated with each measurement. We can then write this as the overlapping SSDU loss \citep[Fig.~5]{yaman_self-supervised_2020} by setting $\A_i^{(k)}=\M_k\A_i,\A_i^{(l)}=\M_l\A_i,\y_i^{(k)}=\M_k\y_i,\y_i^{(l)}=\M_l\y_i$ where $\M_k,\M_l$ are two overlapping mask subsampling sets. Then
\begin{equation*}
    \L_\text{A2A}=\sum_k\sum_{l\neq k}\L_\text{SSDU}^{(k,l)}.
\end{equation*}

\end{proposition}

\paragraph{Learning from invariance} For losses involving equivariant imaging, we draw randomly a group transform at each iteration; for rotation we define the group as $G=\text{SO}(\mathbb{R}^2)$ following \citet{chen_equivariant_2021}.

\paragraph{Data augmentation} For VORTEX \citep{desai_vortex_2022}, noting that the results of their various variants are very similar, for $\T_1$ we use random noise and their random phase errors with $\sigma=\alpha=1$ and for $\T_2$ we use random maximum $\pm 10\%$ shifts and random $\pm 15^{\circ}$ rotations. We also observe reduced performance when adding more transformations including full $\pm 360^{\circ}$ rotation, scaling, or shearing. We do not perform curriculum learning as the original paper does not suggest it significantly affects performance. We observe that when using the VORTEX consistency term on its own without any supervised or MC loss, it learns the trivial $\ft(\y,\A)=\mathbf{0}$, showing the superfluity of VORTEX on our experiments on in-domain data. Note that Noise2Recon \citep{desai_noise2recon_2022} is a special case VORTEX with $\T_2=\mathbf{I}$ and $\T_1$ is random noise.

\paragraph{Adversarial losses} For the adversarial losses, we use the same model $\ft$ as the generator, and the simple convolutional discriminator with skip connections used in \citet{cole_fast_2021}. We perform generator and discriminator steps with ratio 1. Noting that there is a wide range of different GAN flavours, we use the MSE in the adversarial loss calculations following LSGAN \citep{lucic_are_2018}.

\begin{figure}[h]
  \centering
  \includegraphics[width=0.45\textwidth]{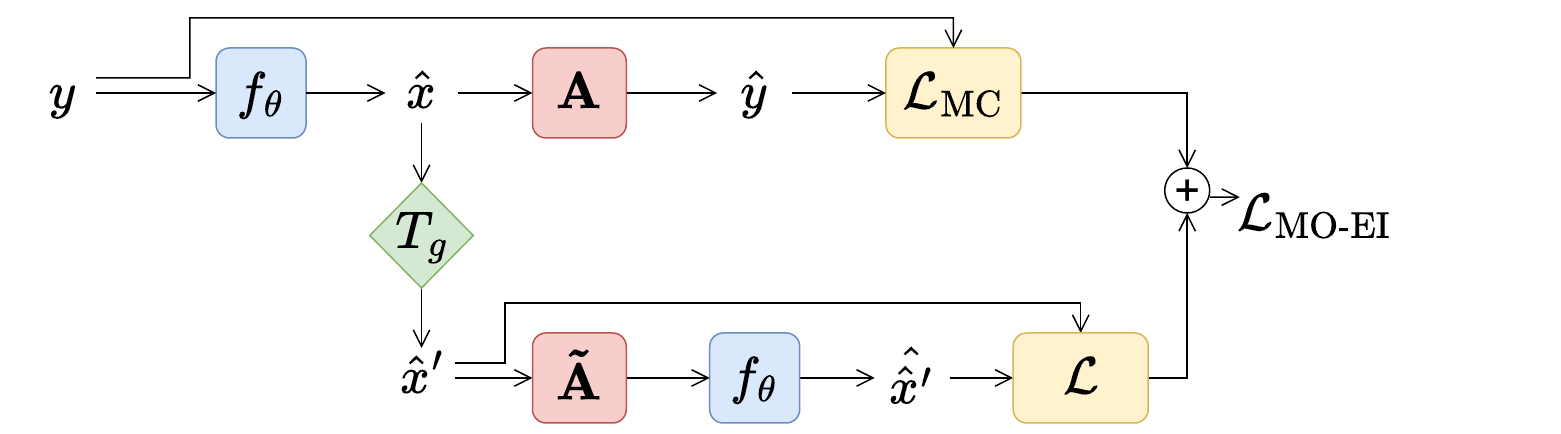}
  \caption{The multi-operator equivariant imaging (MO-EI) loss function.}
  \label{fig:framework}
\end{figure}

\begin{figure}[h]
  \centering
  \includegraphics[width=0.45\textwidth]{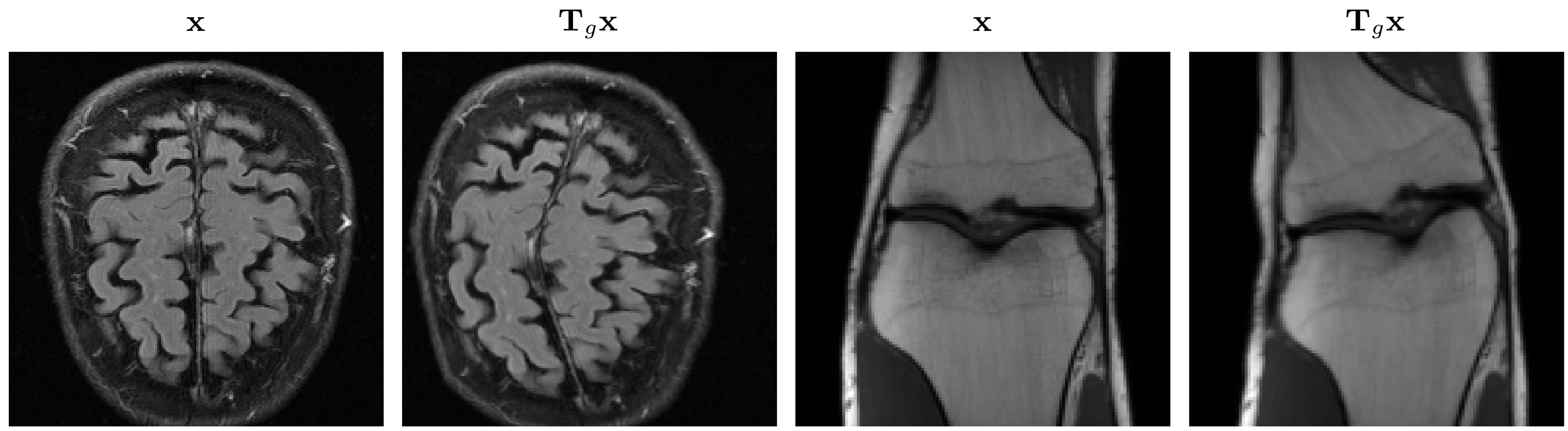}
  \caption{Diffeomorphic transforms.}
  \label{fig:transforms}
\end{figure}

\paragraph{MO-EI} The MO-EI loss function is visualised diagrammatically in \cref{fig:framework}. We let $G=\text{Diffeo}(\mathbb{R}^2)$. As these are very general, we relax the full assumption and instead enforce approximate equivariance to small perturbative distortions by taking the subset of continuous piecewise-affine-based diffeomorphic transforms from \citet{freifeld_transformations_2017,wang_fully_2024}, visualised in \cref{fig:transforms}. For MO-EI, we ablate the choice of transform in \cref{sec:moei_ablation}.

\begin{remark}[MO-EI generalises MOI and EI]
By letting $G$ be the trivial group, we recover MOI \citep{tachella_unsupervised_2022}, and by letting $\mathcal{A}=\{\A\}$, we recover EI \citep{chen_equivariant_2021}.
\end{remark}

\section{Further experimental results}
\label{sec:further_experiments}

\subsection{Statistical significance results}
\label{sec:statsig}
\begin{figure}[h]
\centering
\includegraphics[width=0.4\textwidth]{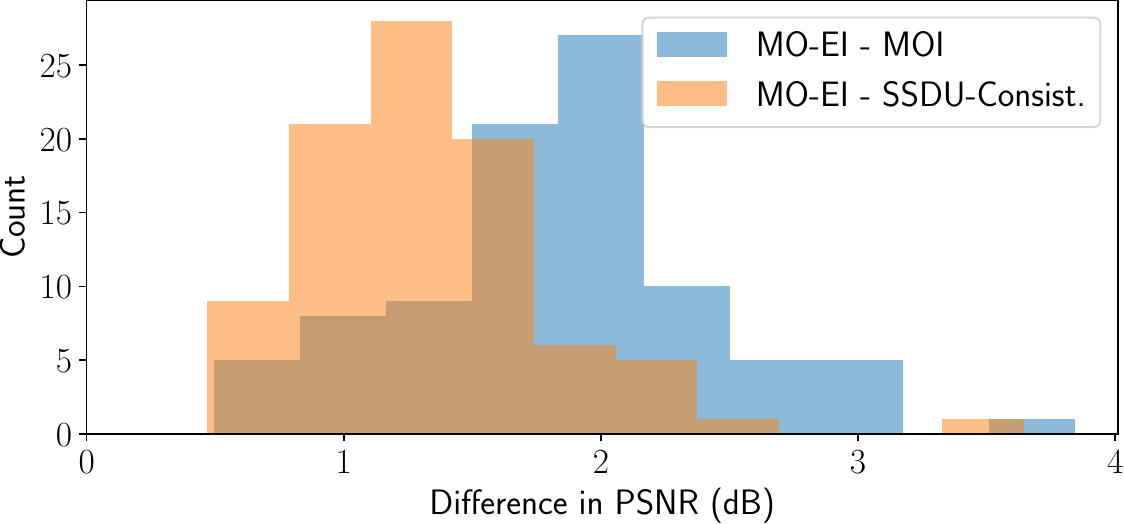}
\caption{Histograms of paired diffs for scen. 1.}
\label{fig:statsignorm}
\end{figure}

To compare the performance of MO-EI with two next-best methods in scenario 1, and likewise for other scenarios, we conduct two statistical tests: paired $t$-test (where normality of the differences is validated in \cref{fig:statsignorm}), and Wilcoxon signed-rank. Results are shown in \cref{tab:statsig}, suggesting, to a 0.5\% significance level, that there is sufficient evidence to reject the null hypothesis and accept the alternative hypothesis that the top-performing method has higher average performance for scenarios 1-5. Note that scenario 7 lacks GT and thus no quantitative metrics were reported.

\begin{table}[h]
\centering
\caption{Statistical significance test p-values comparing test-set PSNR of top method and next-best method(s) for scenarios 1-6.}
\label{tab:statsig}
\resizebox{0.47\textwidth}{!}{
\begin{tabular}{llll}
\hline
Scen. & Comparison & Paired $t$ & Wilcoxon \\ \hline
1 & MO-EI $\leftrightarrow$ MOI              & \SI{3.7e-48}{} & \SI{6.0e-17}{} \\
1 & MO-EI $\leftrightarrow$ SSDU-Consistency & \SI{6.4e-44}{} & \SI{6.0e-17}{} \\
2 & Robust-MO-EI $\leftrightarrow$ Robust-EI & \SI{6.9e-33}{} & \SI{6.0e-17}{} \\
3 & EI $\leftrightarrow$ MOI & \SI{2.5e-26}{} & \SI{8.9e-17}{} \\
4 & Weighted-SSDU $\leftrightarrow$ SSDU-Consistency & \SI{1.9e-11}{} & \SI{1.4e-10}{} \\
5 & Weighted-SSDU $\leftrightarrow$ MOC-SSDU & \SI{3.1e-9}{} & \SI{3.9e-3}{} \\
6 & MOC-t-SSDU $\leftrightarrow$ t-SSDU-Consistency & \SI{5.3e-2}{} & \SI{8.9e-2}{} \\
\hline
\end{tabular}
}
\end{table}

\subsection{Rank of single-coil vs multi-coil operators}
\label{sec:svd}

\begin{figure}[tb]
  \centering
  \includegraphics[width=0.48\textwidth]{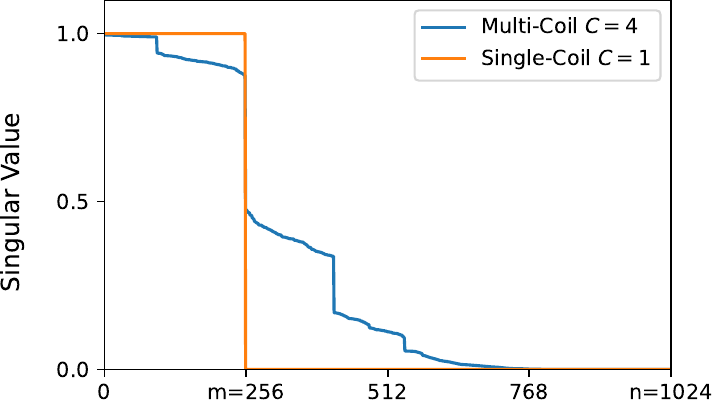}
  \caption{Numerical singular values of single-coil vs multi-coil operators, sorted in descending order.}
  \label{fig:results_svd}
\end{figure}

We further investigate the results of the multi-coil scenario by measuring the size of the null-space compared to the single-coil scenario. We consider a toy $4\times$ accelerated MRI example on $32\times32$ images, so $n=32\times32=1024$, $m=nC/4=256$ for single-coil and $C=4,m=1024$ for multi-coil. We then compute the numerical SVD of the operator $\A$; results are in \cref{fig:results_svd}. The single-coil operator (scenario 1) has exactly $m=\text{rank}(\A)$ non-zero singluar values as expected, whereas the multi-coil operator (scenario 4), now incorporating both masks and sensitivity maps, has many more non-zero singular values, increasing its effective rank \citep{ubaru_fast_2016} and decreasing the size of the ``effective null-space''. For example, consider the case $\sigma=0.1$; the effective $\text{rank}_{\sigma^2}(\A)=686$, so the effective acceleration is $n/m=1.49<4$. This reduces the difficulty of learning in the null-space; we leave for further work tuning the acceleration rate to achieve a similar effective rank as the single-coil case.

\subsection{Scenarios 2, 3, 4 \& 6}
\label{sec:model_ablations}
\begin{figure*}
  \centering
  \includegraphics[width=0.999\textwidth]{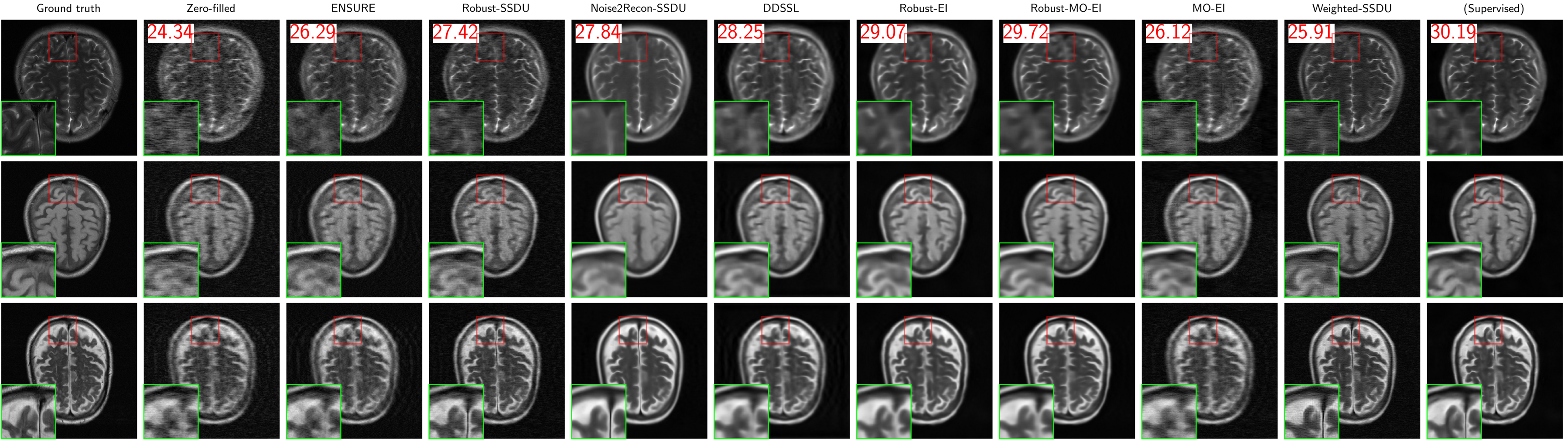}
  \caption{Joint denoise/recon (scenario 2) with test-set PSNR, showing selected highest-performing methods. See \cref{tab:results_scenario2} for more.}
  \label{fig:results_scenario2}
\end{figure*}

\begin{figure*}
    \centering
    \includegraphics[width=0.999\textwidth]{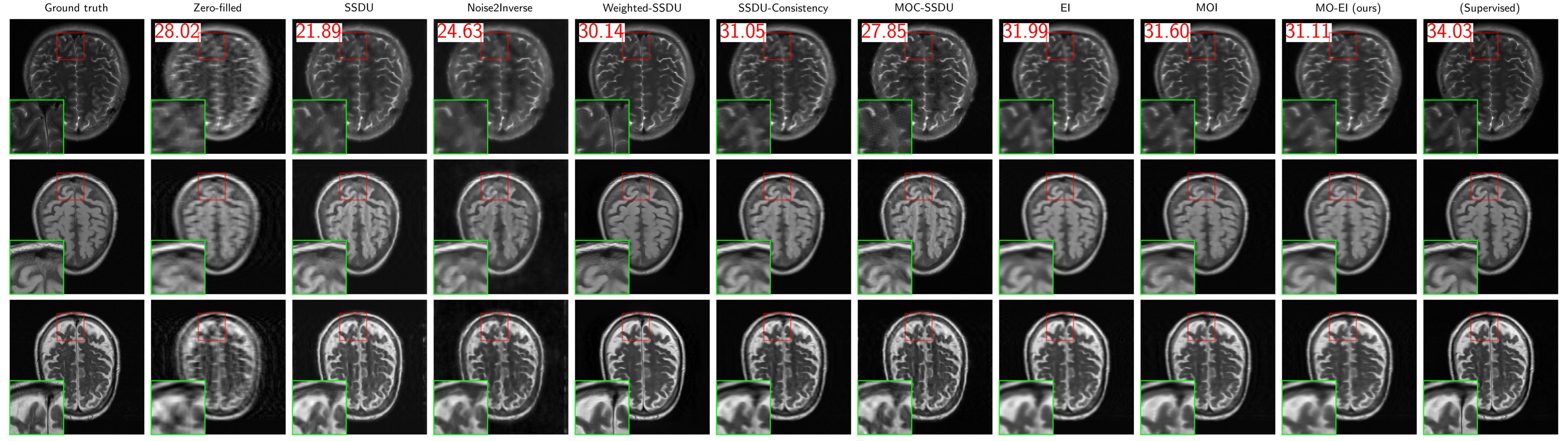}
    \caption{Single-mask (scenario 3) recons with test-set PSNR, showing selected highest-performing methods. See \cref{tab:results_scenario1345} for more.}
    \label{fig:results_scenario3}
\end{figure*}

\begin{figure*}
  \centering
  \includegraphics[width=0.999\textwidth]{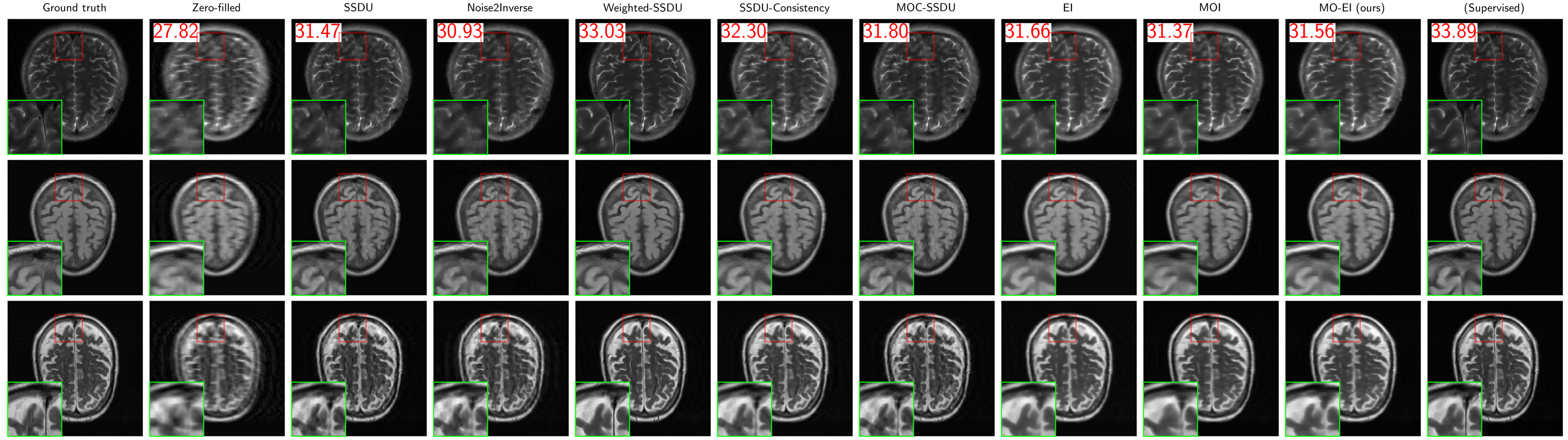}
  \caption{Multicoil brain recons (scenario 4) with test-set PSNR, showing selected highest-performing methods. See \cref{tab:results_scenario1345} for more.}
  \label{fig:results_scenario4}
\end{figure*}

\begin{figure*}
    \centering
    \includegraphics[width=0.999\textwidth]{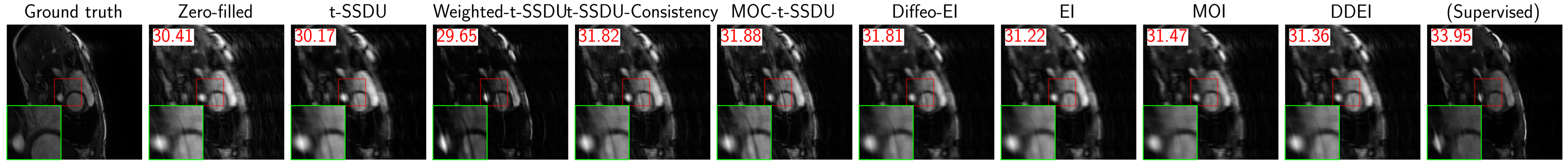}
    \caption{$10\times$ acc. CMRxRecon \citep{wang_cmrxrecon_2023} dynamic cardiac MRI (scenario 6) recons, showing $0$th frame for selected highest-performing methods. See \cref{tab:results_scenario6} for more.}
    \label{fig:results_scenario6}
\end{figure*}

We provide sample reconstructions in \cref{fig:results_scenario2,fig:results_scenario3,fig:results_scenario4,fig:results_scenario6} to complement the quantitative results in \cref{tab:results_scenario1345,tab:results_scenario2,tab:results_scenario6}.

\subsection{Effect of model architecture}

\subsubsection{Unrolled depth}
\label{sec:unroll_depth}

\begin{figure*}[tb]
  \centering
  \includegraphics[width=0.999\textwidth]{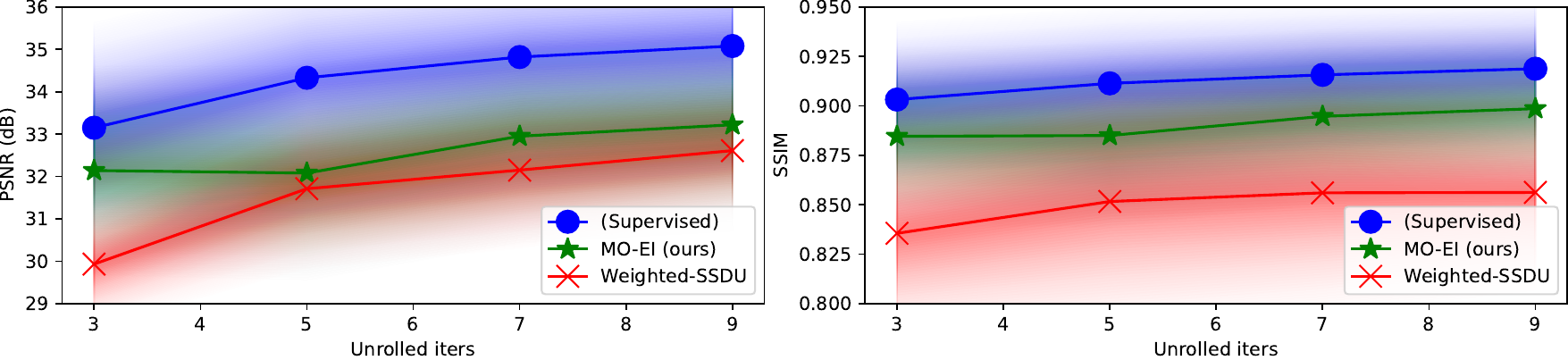}
  \caption{Quantitative results on scenario 1 for Weighted-SSDU, MO-EI, and oracle supervised losses for increasing number of unrolled iterations. Left: PSNR, right: SSIM. Shaded areas repesent one $\sigma$.}
  \label{fig:results_unrolled}
\end{figure*}

We investigate the effect of unrolling depth $u$ (and hence time and memory cost) on performance scaling, starting at $u=3$ (\ie the model on which all results are reported in the main paper) up to $u=9$. Results on scenario 1 are shown in \cref{fig:results_unrolled} on two self-supervised losses and the supervised loss, showing a largely monotonic increase in performance as $u$ increases, suggesting that results presented in the main paper would scale with more expensive models, and that there is a trade-off between model cost and performance.

\subsubsection{Effect of model architecture size}
\label{sec:model_size}

The benchmark results on scenarios 1--4 use an 8.6M parameter model. Although not large by modern standards, this setting may be too large for implementations in constrained hardware configurations. We thus evaluate our benchmark using a model with <1M parameters which would be more suitable for a constrained clinical environment, \eg without extensive GPU compute, by providing a pilot study where we simply swap our benchmark `model` for a 190K parameter unrolled VarNet \citep{hammernik_learning_2018} with a lightweight CNN denoiser \citep{zhang_beyond_2017}. Scenario 1 results in \cref{tab:varnet} show that although the Scenario 1 performance is uniformly decreased, the ranking of the methods is broadly preserved, evidencing the orthogonality of loss choice and model architecture design for this scenario.

\begin{table}[h]
\centering
\caption{Effect of model size: benchmark results on scenario 1 but with a 50$\times$ smaller model (VarNet with flat CNN).}
\label{tab:varnet}
\resizebox{0.47\textwidth}{!}{
\begin{tabular}{llll}
\hline
Loss & PSNR $\uparrow$ & SSIM $\uparrow$ & LPIPS $\downarrow$ \\ \hline
Zero-filled & 27.67\std{2.40} & .7862\std{.05} & .3270\std{.0426} \\
SSDU & 24.24\std{1.50} & .6057\std{.0575} & .3640\std{.0527} \\
Weighted-SSDU & 24.22\std{1.28} & .6237\std{.0761} & .3697\std{.0427} \\
SSDU-Consistency & 27.96\std{2.30} & .7941\std{.0568} & .2801\std{.0500} \\
MOC-SSDU & 26.16\std{1.54} & .6670\std{.0627} & .3597\std{.0432} \\
EI & 29.12\std{2.56} & \textbf{.8349}\std{.0546} & .2714\std{.0380} \\
MOI & 28.68\std{2.55} & .8275\std{.0552} & .2935\std{.0486} \\
MO-EI & \textbf{29.22}\std{2.68} & .8348\std{.0557} & \textbf{.2629}\std{.0493} \\
(Supervised) & 30.86\std{2.50} & .8633\std{.0530} & .1598\std{.0399} \\
\hline
\end{tabular}
}
\end{table}

\subsubsection{Alternative architectures}

While convolutional nets remain popular for imaging tasks \citep{heckel_deep_2024}, attention-based networks have recently demonstrated competing performance \citep{wang_sdum_2025} at the cost of higher computational cost and parameter count due to the self-attention. We swap out the benchmark model for a shallow Restormer \citep{zamir_restormer_2022} architecture and reduce the image size by half to fit the training in GPU memory. Scenario 1 results in \cref{tab:restormer} again show broadly preserved ranking, albeit with a significant performance drop, due to the usage of a shallower network to fit into memory.

\begin{table}[h]
\centering
\caption{Benchmark results with Restormer.}
\label{tab:restormer}
\resizebox{0.43\textwidth}{!}{ 
\begin{tabular}{llll}
\hline
Loss & PSNR $\uparrow$ & SSIM $\uparrow$ & LPIPS $\downarrow$ \\ \hline
Zero-filled & 25.17\std{2.22} & .7381\std{.05} & .1956\std{.0398} \\
MC & 25.16\std{2.22} & .7373\std{.0490} & .1941\std{.0400} \\
SSDU & 25.18\std{1.15} & .7503\std{.0511} & .1715\std{.0345} \\
Noise2Inverse & 25.22\std{3.17} & .7520\std{.0823} & .1870\std{.0450} \\
Weighted-SSDU & 25.48\std{1.10} & .7789\std{.0450} & .1450\std{.0359} \\
SSDU-Consistency & 24.92\std{1.66} & .8070\std{.0368} & .1316\std{.0316} \\
MOC-SSDU & 23.28\std{1.74} & .7301\std{.0473} & .1181\std{.0301} \\
UAIR & 0.90\std{1.20} & .0921\std{.0214} & .2886\std{.0429} \\
VORTEX & 25.06\std{2.19} & .7282\std{.0472} & .1897\std{.0397} \\
EI & 27.15\std{2.73} & .8180\std{.0434} & .1213\std{.0366} \\
MOI & 27.43\std{2.63} & .8358\std{.0399} & .1438\std{.0434} \\
MO-EI & 27.22\std{2.80} & .8319\std{.0432} & .1307\std{.0392} \\
\hline
\end{tabular}
}
\end{table}

\subsection{Pretrained diffusion model results}
\label{sec:diffusion}
\input{sec/results_table_diffusion}

We evidence the extensibility of our framework by including pilot results in \cref{tab:results_diffusion} on SotA diffusion methods \citep{chung_diffusion_2022} using recently proposed GT-free pretrained models \citep{aali_ambient_2024}. Note that it is infeasible to train these diffusion models \citep{karras_elucidating_2022} on the modest-size dataset used in any of the scenarios considered in the main paper. Therefore, these results cannot be directly compared to other methods from the main paper, since the pretrained models from \citep{aali_ambient_2024} are trained on datasets of much larger size (${\sim}10^4$ images), perhaps covering our test images. We apply the diffusion methods to scenarios 1-4 in \cref{tab:results_diffusion}. Since diffusion methods sample a point estimate from the posterior, it is expected that they achieve good perceptual performance, but poorer distortion metrics compared to MMSE estimators \citep{blau_perception-distortion_2018}. In the in-distribution setting, the one-step MMSE reconstruction improves the distortion result at the expense of perceptual performance, mirroring the end-to-end loss performances. The noisy performance suggests that the model struggles to generalise to this setting. DPS and ALD methods require a large number of iterations, so they are sampled using 500 denoising iterations (NFEs) per image \citep{aali_ambient_2024}.

\subsection{Ablations}
\subsubsection{Ablation of MO-EI loss}
\label{sec:moei_ablation}
The components of our proposed loss function are ablated by interpolating between existing methods MOI and EI and the proposed method MO-EI, and report results in \cref{tab:moei_ablation}. We note that the base MOI and EI have similar results, and adding the two independent components of a) MO-EI and b) the diffeomorphic transforms helps, but combining these two provides the best performance in our proposed method, as per our theoretical framework.

\begin{table}[t]
\centering
\caption{Ablation of components of the proposed MO-EI loss function.}
\label{tab:moei_ablation}
\resizebox{0.43\textwidth}{!}{
\begin{tabular}{llll}
\hline
Loss & PSNR $\uparrow$ & SSIM $\uparrow$ \\ \hline
MOI & 30.29\std{2.88} & .8651\std{.0528} \\
EI (rotate) & 30.26\std{2.61} & .8523\std{.0542} \\
EI (diffeo) & 31.26\std{2.88} & .8741\std{.0522} \\
MO-EI (rotate) & 30.62\std{2.70} & .8575\std{.0553} \\
\textbf{MO-EI (diffeo)} & \textbf{32.14}\std{2.73} & \textbf{.8846}\std{.0498} \\
\hline
\end{tabular}
}
\end{table}

\subsubsection{Ablation of SSDU loss}
\label{sec:ssdu_ablation}
We ablate for the components of SSDU \citep{yaman_multi-mask_2022} and weighted-SSDU \citep{millard_theoretical_2023} and report results in \cref{tab:ssdu_ablation}. While \citep{yaman_multi-mask_2022} construct variable-density 2D partitioning masks with an ACS block, we were only able to achieve good performance using uniform-density partitioning (\ie Bernoulli) without an ACS block. This aligns with theory, since the lack of ACS in $\M_1$ means that both $\M_1$ and $\M_2$ contain low-frequency lines; otherwise, $\y_2$ would contain almost no low-frequency image content and thus diverge the loss. We reproduce the weighted-SSDU results reported in \citep{millard_theoretical_2023} reflecting their theory: results improve when adding the weighting even when using the original 2D partitioning, and improves further when using 1D partitioning masks, since these $\M_1$ are then drawn from the same original distribution of $\M$, but only if these also do not have ACS lines.

\begin{table}[t]
\centering
\caption{Ablation of components of SSDU \citep{yaman_multi-mask_2022} and weighted-SSDU \citep{millard_theoretical_2023} loss functions. * = methods reported in main paper.}
\label{tab:ssdu_ablation}
\resizebox{0.47\textwidth}{!}{
\begin{tabular}{llll}
\hline
Loss & PSNR $\uparrow$ & SSIM $\uparrow$ \\ \hline
SSDU (uniform-density no ACS)* & 27.98\std{1.43} & .7485\std{.0667} \\
SSDU (variable-density with ACS) & 18.27\std{1.01} & .4876\std{.0598} \\
\textbf{Weighted-SSDU (1D partition)}* & \textbf{29.93}\std{1.66} & \textbf{.8355}\std{.0626} \\
Weighted-SSDU (2D partition) & 28.62\std{1.59} & .7489\std{.0692} \\
Weighted-SSDU (1D with ACS) & 13.41\std{.63} & .3875\std{.0479} \\
SSDU (1D partition) & 27.38\std{1.87} & .8055\std{.0658} \\
\hline
\end{tabular}
}
\end{table}

\subsection{Transfer of benchmark to different imaging modalities}
\label{sec:hsi}
\subsubsection{Hyperspectral imaging}
We demonstrate the transferability of our modular benchmark on a different scientific imaging modality where GT images do not exist. Hyperspectral imaging aims to obtain high quality multi-channel images of the Earth for remote sensing and environmental monitoring, but observations may be degraded due to impulse noise arising from missing lines and thermal noise \citep{pang_hir-diff_2024,bodrito_trainable_2021,sidorov_deep_2019}. Hyperspectral restoration is therefore an important preprocessing task, but GT is unavailable \citep{bodrito_trainable_2021,sidorov_deep_2019} because of the non-stationary nature of satellites. 

Our modular benchmark framework facilitates evaluating SSI methods on any imaging inverse problem such as the hyperspectral restoration problem, and we demonstrate a proof-of-concept showing the ease of adapting the benchmark to a different problem domain. Concretely, we simply swap out the following modules:

\begin{enumerate}
    \item \lstinline[basicstyle=\normalsize\ttfamily]$physics = deepinv.physics.MRI()$ $\rightarrow$
    \lstinline[basicstyle=\normalsize\ttfamily]$deepinv.physics.Inpainting(noise_model=GaussianNoise())$, \ie random inpainting with 1D masks (50\% dropout) with additive noise ($\sigma=0.1$);
    \item \lstinline[basicstyle=\normalsize\ttfamily]$dataset = FastMRISliceDataset(...)$ $\rightarrow$ \lstinline[basicstyle=\normalsize\ttfamily]$NBUDataset(...)$, where we use a subset of 150 $256\times256\times4$ WorldView-2 satellite patches provided by \citet{meng_large-scale_2021}.
\end{enumerate}

\begin{figure*}
  \centering
  \includegraphics[width=0.9\textwidth]{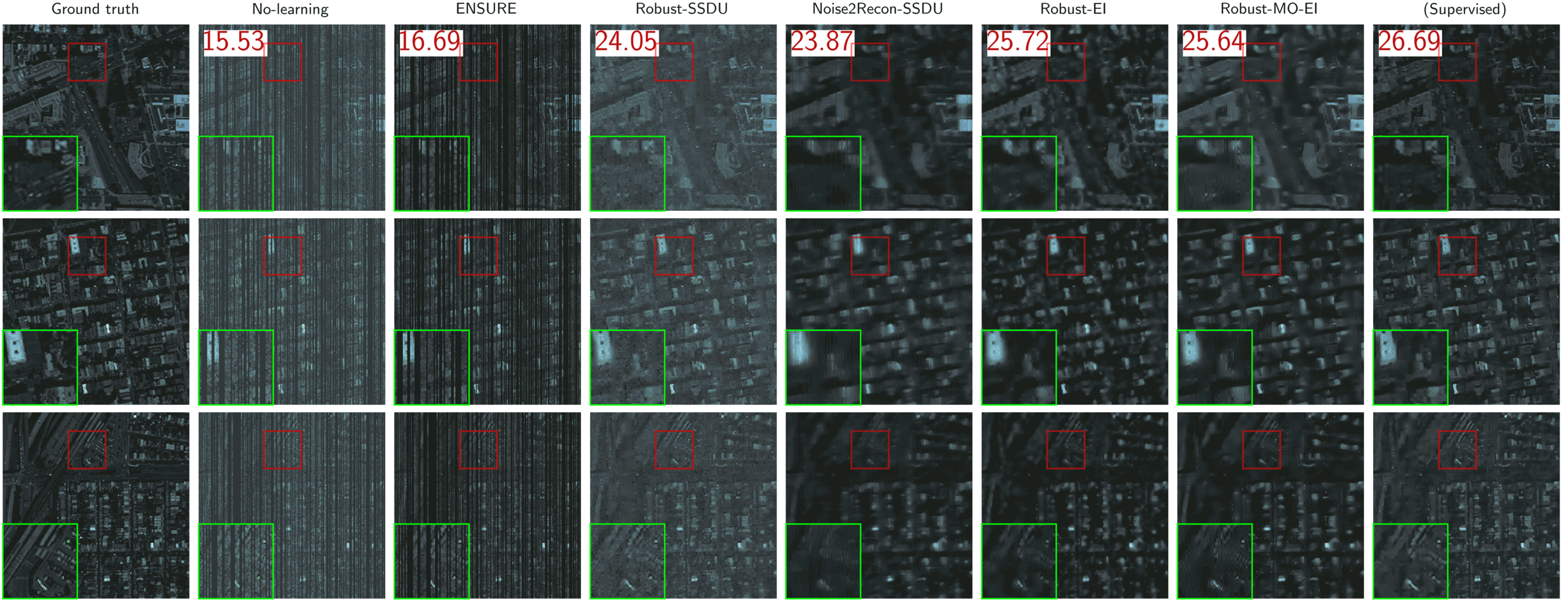}
  \caption{Sample reconstructions for hyperspectral image restoration.}
  \label{fig:test_hsi}
\end{figure*}

Qualitative and quantitative results are shown in \cref{fig:test_hsi,tab:test_hsi}, benchmarking self-supervised losses for joint reconstruction and denoising, with the no-learning result showing the input since $\A^\top\y=\y$. The results show the superiority of Robust-EI, with ENSURE failing to recover information in the null-space, and Robust-SSDU and Noise2Recon-SSDU less able to produce spatially and spectrally faithful reconstructions. We conclude that Robust-MO-EI performs less well due to the inappropriate application of diffeomorphic invariance to urban satellite images, where straight lines should remain straight.

\begin{table}[t]
\centering
\caption{Test-set results for hyperspectral image restoration.}
\label{tab:test_hsi}
\resizebox{0.45\textwidth}{!}{
\begin{tabular}{llll}
\hline
Loss & PSNR $\uparrow$ & SSIM $\uparrow$ \\ \hline
Zero-filled & 15.53\std{.59} & .1435\std{.01} \\
ENSURE & 16.69\std{.75} & .2074\std{.0155} \\
Robust-SSDU & 24.05\std{.31} & .4965\std{.0236} \\
Noise2Recon-SSDU & 23.87\std{.75} & .5216\std{.0366} \\
\textbf{Robust-EI} & \textbf{25.72}\std{.72} & \textbf{.6601}\std{.0243} \\
Robust-MO-EI & 25.64\std{.72} & .6535\std{.0242} \\
(Supervised) & 26.69\std{.70} & .7282\std{.0201} \\
\hline
\end{tabular}
}
\end{table}

\subsubsection{Sparse-view CT}

We also perform the same transferability experiment as above but on a different medical task, sparse-view computed tomography reconstruction, where, similar to MRI, one acquires fewer projections to reduce the acquisition time, leading to a worse image quality \citep{sidky_report_2022}. We consider a 45 angle, 2D tomography problem analogous to the benchmark Scenario 1, where the forward operator is a subsampled Radon transform. We train and test the benchmarked losses on a FBPConvNet-style architecture where the input to the U-Net is the filtered back-projection \citep{jin_deep_2017}, and simulate measurements from the CT100 dataset, a dataset of 100 in-vivo CT images from the cancer imaging archive \citep{clark_cancer_2013}. The results in \cref{tab:test_ct} show a wide spread in performance, with similar rankings to that in accelerated MRI.

\begin{table}[t]
\centering
\caption{Test-set results for sparse-view CT.}
\label{tab:test_ct}
\resizebox{0.43\textwidth}{!}{ 
\begin{tabular}{llll}
\hline
Loss & PSNR $\uparrow$ & SSIM $\uparrow$ & LPIPS $\downarrow$ \\ \hline
Backprojection & 17.46\std{2.19} & .4283\std{.11} & .2171\std{.0436} \\
SSDU & 34.70\std{2.56} & .8890\std{.0344} & .1277\std{.0417} \\
Noise2Inverse & 30.07\std{3.00} & .8215\std{.0556} & .2334\std{.0570} \\
SSDU-Consistency & 34.85\std{2.69} & .8937\std{.0349} & .1292\std{.0421} \\
MOC-SSDU & 34.87\std{2.78} & .8899\std{.0378} & .1320\std{.0436} \\
Adversarial & 21.60\std{.17} & .3252\std{.2280} & .1322\std{.0437} \\
UAIR & 22.57\std{2.47} & .8478\std{.0351} & .1448\std{.0484} \\
VORTEX & nan\std{.00} & nan\std{.0000} & nan\std{.0000} \\
EI & 34.88\std{2.78} & .8897\std{.0378} & .1320\std{.0437} \\
MOI & 34.88\std{2.78} & .8898\std{.0378} & .1321\std{.0436} \\
MO-EI & 34.89\std{2.78} & .8902\std{.0376} & .1320\std{.0435} \\
\hline
\end{tabular}
}
\end{table}

%% file: sec/results_table_diffusion.tex
\begin{table*}[!h]
\centering
\caption{Pilot results on diffusion sampling methods with models pretrained without ground truth, applied to benchmark scenarios. Note that the results cannot be directly compared to tables from the main paper, as these models, while pretrained on measurement only data, required substantially larger datasets for training. DPS = diffusion posterior sampling \citep{chung_diffusion_2022,aali_ambient_2024}, ALD = annealed Langevin dynamics \citep{jalal_robust_2021,aali_ambient_2024}.}
\label{tab:results_diffusion}
\resizebox{\textwidth}{!}{
\begin{tabular}{lccc|ccc|ccc|ccc}
\hline
\multicolumn{1}{r}{{Scenario}} &
\multicolumn{3}{c}{\textbf{(1)} Brain MRI 6$\times$ acc.} &
\multicolumn{3}{c}{\textbf{(2)} Noisy $\sigma=.1$} &
\multicolumn{3}{c}{\textbf{(3)} Single-mask} &
\multicolumn{3}{c}{\textbf{(4)} Multicoil $C=4$} \\
\cline{2-13}
Loss & PSNR$\uparrow$ & SSIM$\uparrow$ & LPIPS$\downarrow$
 & PSNR$\uparrow$ & SSIM$\uparrow$ & LPIPS$\downarrow$
 & PSNR$\uparrow$ & SSIM$\uparrow$ & LPIPS$\downarrow$
 & PSNR$\uparrow$ & SSIM$\uparrow$ & LPIPS$\downarrow$ \\ \hline

Zero-filled & 27.65\std{2.40} & 0.7857\std{0.0533} & 0.3270\std{0.0471} & 24.38\std{1.03} & 0.4438\std{0.0348} & 0.4636\std{0.0517} & 28.23\std{2.29} & 0.7951\std{0.0495} & 0.3494\std{0.0491} & 27.81\std{2.40} & 0.7984\std{0.0511} & 0.3175\std{0.0475} \\
Ambient-DPS & 29.49\std{1.52} & 0.7368\std{0.0504} & 0.1726\std{0.0395} & \textbf{26.10}\std{0.98} & \textbf{0.5421}\std{0.0483} & 0.3306\std{0.0535} & 29.27\std{1.54} & 0.7272\std{0.0512} & 0.1792\std{0.0407} & 29.98\std{1.66} & 0.7552\std{0.0506} & \textbf{0.1532}\std{0.0361} \\
Ambient-ALD & 29.52\std{1.60} & 0.7367\std{0.0518} & \textbf{0.1707}\std{0.0376} & \textbf{26.10}\std{0.98} & 0.5420\std{0.0472} & \textbf{0.3300}\std{0.0548} & 29.38\std{1.54} & 0.7319\std{0.0513} & \textbf{0.1753}\std{0.0409} & 29.97\std{1.60} & 0.7550\std{0.0491} & 0.1538\std{0.0349} \\
Ambient-One-Step & \textbf{31.34}\std{2.29} & \textbf{0.8673}\std{0.0480} & 0.1743\std{0.0459} & 24.93\std{0.96} & 0.4655\std{0.0372} & 0.4303\std{0.0547} & \textbf{31.51}\std{2.17} & \textbf{0.8686}\std{0.0467} & 0.2000\std{0.0509} & \textbf{31.53}\std{2.21} & \textbf{0.8741}\std{0.0459} & 0.1685\std{0.0461} \\

\hline
\end{tabular}
}
\end{table*}

%% file: references.bib
@inproceedings{sechaud_equivariant_2025,
	title = {Equivariant {Splitting}: {Self}-supervised learning from incomplete data},
	shorttitle = {Equivariant {Splitting}},
	language = {en},
	urldate = {2026-05-07},
	booktitle = {Proceedings of {The} {Fourteenth} {International} {Conference} on {Learning} {Representations}},
	author = {Sechaud, Victor and Scanvic, Jérémy and Barthélemy, Quentin and Abry, Patrice and Tachella, Julián},
	month = oct,
	year = {2025},
}

@misc{zbontar_fastmri_2018,
	title = {{fastMRI}: {An} {Open} {Dataset} and {Benchmarks} for {Accelerated} {MRI}},
	shorttitle = {{fastMRI}},
	language = {en},
	urldate = {2024-09-06},
	journal = {arXiv.org},
	author = {Zbontar, Jure and Knoll, Florian and Sriram, Anuroop and Murrell, Tullie and Huang, Zhengnan and Muckley, Matthew J. and Defazio, Aaron and Stern, Ruben and Johnson, Patricia and Bruno, Mary and Parente, Marc and Geras, Krzysztof J. and Katsnelson, Joe and Chandarana, Hersh and Zhang, Zizhao and Drozdzal, Michal and Romero, Adriana and Rabbat, Michael and Vincent, Pascal and Yakubova, Nafissa and Pinkerton, James and Wang, Duo and Owens, Erich and Zitnick, C. Lawrence and Recht, Michael P. and Sodickson, Daniel K. and Lui, Yvonne W.},
	month = nov,
	year = {2018},
}

@misc{wang_sdum_2025,
	title = {{SDUM}: {A} {Scalable} {Deep} {Unrolled} {Model} for {Universal} {MRI} {Reconstruction}},
	shorttitle = {{SDUM}},
	language = {en},
	urldate = {2026-05-07},
	author = {Wang, Puyang and Guo, Pengfei and Chai, Keyi and Zhou, Jinyuan and Xu, Daguang and Jiang, Shanshan},
	month = dec,
	year = {2025},
}

@misc{schut_equivariance2inverse_2025,
	title = {{Equivariance2Inverse}: {A} {Practical} {Self}-{Supervised} {CT} {Reconstruction} {Method} {Benchmarked} on {Real}, {Limited}-{Angle}, and {Blurred} {Data}},
	shorttitle = {{Equivariance2Inverse}},
	doi = {10.1109/TCI.2026.3684416},
	language = {en},
	urldate = {2026-05-06},
	journal = {arXiv.org},
	author = {Schut, Dirk Elias and Graas, Adriaan and van Liere, Robert and van Leeuwen, Tristan},
	month = oct,
	year = {2025},
}

@article{jin_deep_2017,
	title = {Deep {Convolutional} {Neural} {Network} for {Inverse} {Problems} in {Imaging}},
	volume = {26},
	doi = {10.1109/TIP.2017.2713099},
	number = {9},
	urldate = {2026-05-07},
	journal = {IEEE Transactions on Image Processing},
	author = {Jin, Kyong Hwan and McCann, Michael T. and Froustey, Emmanuel and Unser, Michael},
	month = sep,
	year = {2017},
	pages = {4509--4522},
}

@article{clark_cancer_2013,
	title = {The {Cancer} {Imaging} {Archive} ({TCIA}): {Maintaining} and {Operating} a {Public} {Information} {Repository}},
	volume = {26},
	doi = {10.1007/s10278-013-9622-7},
	language = {en},
	number = {6},
	urldate = {2026-05-07},
	journal = {Journal of Digital Imaging},
	author = {Clark, Kenneth and Vendt, Bruce and Smith, Kirk and Freymann, John and Kirby, Justin and Koppel, Paul and Moore, Stephen and Phillips, Stanley and Maffitt, David and Pringle, Michael and Tarbox, Lawrence and Prior, Fred},
	month = dec,
	year = {2013},
	pages = {1045--1057},
}

@article{uecker_espiriteigenvalue_2014,
	title = {{ESPIRiT}—an eigenvalue approach to autocalibrating parallel {MRI}: {Where} {SENSE} meets {GRAPPA}},
	volume = {71},
	doi = {10.1002/mrm.24751},
	language = {en},
	number = {3},
	urldate = {2026-02-23},
	journal = {Magnetic Resonance in Medicine},
	author = {Uecker, Martin and Lai, Peng and Murphy, Mark J. and Virtue, Patrick and Elad, Michael and Pauly, John M. and Vasanawala, Shreyas S. and Lustig, Michael},
	year = {2014},
	pages = {990--1001},
}

@inproceedings{wang_perspective-equivariance_2025,
	address = {Cham},
	title = {Perspective-{Equivariance} for {Unsupervised} {Imaging} with {Camera} {Geometry}},
	isbn = {978-3-031-91585-7},
	doi = {10.1007/978-3-031-91585-7_8},
	language = {en},
	booktitle = {Computer {Vision} – {ECCV} 2024 {Workshops}},
	publisher = {Springer Nature Switzerland},
	author = {Wang, Andrew and Davies, Mike},
	editor = {Del Bue, Alessio and Canton, Cristian and Pont-Tuset, Jordi and Tommasi, Tatiana},
	year = {2025},
	pages = {116--134},
}

@article{tachella_self-supervised_2026,
	title = {Self-supervised learning from noisy and incomplete data},
	volume = {20},
	issn = {1932-8346},
	url = {https://doi.org/10.1108/FTSIG-10-2025-0133},
	doi = {10.1108/FTSIG-10-2025-0133},
	number = {2},
	urldate = {2026-06-24},
	journal = {Foundations and Trends in Signal Processing},
	author = {Tachella, Julián and Davies, Mike},
	month = apr,
	year = {2026},
	pages = {85--184},
}

@inproceedings{terris_reconstruct_2025,
	title = {Reconstruct {Anything} {Model} a lightweight general model for computational imaging},
	booktitle = {Proceedings of {The} {Fourteenth} {International} {Conference} on {Learning} {Representations}},
	author = {Terris, Matthieu and Hurault, Samuel and Song, Maxime and Tachella, Julián},
	month = oct,
	year = {2025},
}

@article{sidky_report_2022,
	title = {Report on the {AAPM} deep-learning sparse-view {CT} grand challenge},
	volume = {49},
	doi = {10.1002/mp.15489},
	language = {en},
	number = {8},
	urldate = {2026-05-07},
	journal = {Medical Physics},
	author = {Sidky, Emil Y. and Pan, Xiaochuan},
	year = {2022},
	pages = {4935--4943},
}

@article{kiss_benchmarking_2025,
	title = {Benchmarking learned algorithms for computed tomography image reconstruction tasks},
	volume = {3},
	doi = {10.3934/ammc.2025001},
	language = {en},
	number = {0},
	urldate = {2026-04-18},
	journal = {Applied Mathematics for Modern Challenges},
	publisher = {Applied Mathematics for Modern Challenges},
	author = {Kiss, Maximilian B. and Biguri, Ander and Shumaylov, Zakhar and Sherry, Ferdia and Batenburg, K. Joost and Schönlieb, Carola-Bibiane and Lucka, Felix},
	month = feb,
	year = {2025},
	pages = {1--43},
}

@inproceedings{zamir_restormer_2022,
	title = {Restormer: {Efficient} {Transformer} for {High}-{Resolution} {Image} {Restoration}},
	urldate = {2026-05-07},
	booktitle = {2022 {IEEE}/{CVF} {Conference} on {Computer} {Vision} and {Pattern} {Recognition} ({CVPR})},
	author = {Zamir, Syed Waqas and Arora, Aditya and Khan, Salman and Hayat, Munawar and Khan, Fahad Shahbaz and Yang, Ming–Hsuan},
	month = jun,
	year = {2022},
	pages = {5718--5729},
}

@misc{gruber_sparse2inverse_2024,
	title = {{Sparse2Inverse}: {Self}-supervised inversion of sparse-view {CT} data},
	shorttitle = {{Sparse2Inverse}},
	url = {http://arxiv.org/abs/2402.16921},
	doi = {10.48550/arXiv.2402.16921},
	urldate = {2026-05-06},
	publisher = {arXiv},
	author = {Gruber, Nadja and Schwab, Johannes and Gizewski, Elke and Haltmeier, Markus},
	month = feb,
	year = {2024},
	note = {arXiv:2402.16921 [eess]},
}

@misc{xu_towards_2026,
	title = {Towards a {Unified} {Theoretical} {Framework} for {Self}-{Supervised} {MRI} {Reconstruction}},
	url = {http://arxiv.org/abs/2601.04775},
	doi = {10.48550/arXiv.2601.04775},
	urldate = {2026-02-24},
	publisher = {arXiv},
	author = {Xu, Siying and Hammernik, Kerstin and Rueckert, Daniel and Gatidis, Sergios and Küstner, Thomas},
	month = feb,
	year = {2026},
	note = {arXiv:2601.04775 [eess]},
}

@misc{zheng_inversebench_2025,
	title = {{InverseBench}: {Benchmarking} {Plug}-and-{Play} {Diffusion} {Priors} for {Inverse} {Problems} in {Physical} {Sciences}},
	shorttitle = {{InverseBench}},
	url = {http://arxiv.org/abs/2503.11043},
	doi = {10.48550/arXiv.2503.11043},
	urldate = {2026-02-24},
	publisher = {arXiv},
	author = {Zheng, Hongkai and Chu, Wenda and Zhang, Bingliang and Wu, Zihui and Wang, Austin and Feng, Berthy T. and Zou, Caifeng and Sun, Yu and Kovachki, Nikola and Ross, Zachary E. and Bouman, Katherine L. and Yue, Yisong},
	month = sep,
	year = {2025},
	note = {arXiv:2503.11043 [cs]},
}

@article{tachella_deepinverse_2025,
	title = {{DeepInverse}: {A} {Python} package for solving imaging inverse problems with deep learning},
	volume = {10},
	issn = {2475-9066},
	shorttitle = {{DeepInverse}},
	url = {https://joss.theoj.org/papers/10.21105/joss.08923},
	doi = {10.21105/joss.08923},
	language = {en},
	number = {115},
	urldate = {2026-02-22},
	journal = {Journal of Open Source Software},
	author = {Tachella, Julián and Terris, Matthieu and Hurault, Samuel and Wang, Andrew and Davy, Leo and Scanvic, Jérémy and Sechaud, Victor and Vo, Romain and Moreau, Thomas and Davies, Thomas and Chen, Dongdong and Laurent, Nils and Monroy, Brayan and Dong, Jonathan and Hu, Zhiyuan and Nguyen, Minh-Hai and Sarron, Florian and Weiss, Pierre and Escande, Paul and Massias, Mathurin and Modrzyk, Thibaut and Levac, Brett and Liaudat, Tobías I. and Song, Maxime and Hertrich, Johannes and Neumayer, Sebastian and Schramm, Georg},
	month = nov,
	year = {2025},
	pages = {8923},
}

@article{zhang_beyond_2017,
	title = {Beyond a {Gaussian} {Denoiser}: {Residual} {Learning} of {Deep} {CNN} for {Image} {Denoising}},
	volume = {26},
	number = {7},
	journal = {IEEE Transactions on Image Processing},
	author = {Zhang, Kai and Zuo, Wangmeng and Chen, Yunjin and Meng, Deyu and Zhang, Lei},
	month = jul,
	year = {2017},
	pages = {3142--3155},
}

@article{karras_elucidating_2022,
	title = {Elucidating the {Design} {Space} of {Diffusion}-{Based} {Generative} {Models}},
	volume = {35},
	journal = {Advances in Neural Information Processing Systems},
	author = {Karras, Tero and Aittala, Miika and Aila, Timo and Laine, Samuli},
	month = dec,
	year = {2022},
	pages = {26565--26577},
}

@inproceedings{blau_perception-distortion_2018,
	title = {The {Perception}-{Distortion} {Tradeoff}},
	booktitle = {2018 {IEEE}/{CVF} {Conference} on {Computer} {Vision} and {Pattern} {Recognition}},
	author = {Blau, Yochai and Michaeli, Tomer},
	month = jun,
	year = {2018},
}

@article{yu_validation_2022,
	title = {Validation and {Generalizability} of {Self}-{Supervised} {Image} {Reconstruction} {Methods} for {Undersampled} {MRI}},
	volume = {1},
	number = {September 2022 issue},
	journal = {Machine Learning for Biomedical Imaging},
	author = {Yu, Thomas and Hilbert, Tom and Piredda, Gian Franco and Joseph, Arun and Bonanno, Gabriele and Zenkhri, Salim and Omoumi, Patrick and Cuadra, Meritxell Bach and Canales Rodriguez, Erick and Kober, Tobias and Thiran, Jean-Philippe},
	month = sep,
	year = {2022},
	pages = {1--31},
}

@article{ying_joint_2007,
	title = {Joint image reconstruction and sensitivity estimation in {SENSE} ({JSENSE})},
	volume = {57},
	number = {6},
	journal = {Magnetic Resonance in Medicine},
	author = {Ying, Leslie and Sheng, Jinhua},
	year = {2007},
	pages = {1196--1202},
}

@inproceedings{zhang_unreasonable_2018,
	title = {The {Unreasonable} {Effectiveness} of {Deep} {Features} as a {Perceptual} {Metric}},
	doi = {10.1109/CVPR.2018.00068},
	booktitle = {2018 {IEEE}/{CVF} {Conference} on {Computer} {Vision} and {Pattern} {Recognition}},
	author = {Zhang, Richard and Isola, Phillip and Efros, Alexei A. and Shechtman, Eli and Wang, Oliver},
	month = jun,
	year = {2018},
}

@inproceedings{jalal_robust_2021,
	title = {Robust {Compressed} {Sensing} {MRI} with {Deep} {Generative} {Priors}},
	booktitle = {Advances in {Neural} {Information} {Processing} {Systems}},
	author = {Jalal, Ajil and Arvinte, Marius and Daras, Giannis and Price, Eric and Dimakis, Alexandros G and Tamir, Jon},
	year = {2021},
}

@inproceedings{chung_diffusion_2022,
	title = {Diffusion {Posterior} {Sampling} for {General} {Noisy} {Inverse} {Problems}},
	booktitle = {The {Eleventh} {International} {Conference} on {Learning} {Representations}},
	author = {Chung, Hyungjin and Kim, Jeongsol and Mccann, Michael Thompson and Klasky, Marc Louis and Ye, Jong Chul},
	month = sep,
	year = {2022},
}

@inproceedings{joo_aespa_2025,
	title = {{AeSPa} : {Attention}-guided {Self}-supervised {Parallel} {Imaging} for {MRI} {Reconstruction}},
	booktitle = {Proceedings of the {IEEE}/{CVF} {Conference} on {Computer} {Vision} and {Pattern} {Recognition}},
	author = {Joo, Jinho and Kim, Hyeseong and Won, Hyeyeon and Lee, Deukhee and Eo, Taejoon and Hwang, Dosik},
	year = {2025},
}

@article{xu_self-supervised_2025,
	title = {Self-{Supervised} {Feature} {Learning} for {Cardiac} {Cine} {MR} {Image} {Reconstruction}},
	volume = {44},
	number = {9},
	journal = {IEEE Transactions on Medical Imaging},
	author = {Xu, Siying and Früh, Marcel and Hammernik, Kerstin and Lingg, Andreas and Kübler, Jens and Krumm, Patrick and Rueckert, Daniel and Gatidis, Sergios and Küstner, Thomas},
	year = {2025},
}

@inproceedings{qin_ground-truth_2023,
	title = {Ground-{Truth} {Free} {Meta}-{Learning} for {Deep} {Compressive} {Sampling}},
	language = {en},
	booktitle = {Proceedings of the {IEEE}/{CVF} {Conference} on {Computer} {Vision} and {Pattern} {Recognition}},
	author = {Qin, Xinran and Quan, Yuhui and Pang, Tongyao and Ji, Hui},
	year = {2023},
	pages = {9947--9956},
}

@inproceedings{pang_recorrupted--recorrupted_2021,
	title = {Recorrupted-to-{Recorrupted}: {Unsupervised} {Deep} {Learning} for {Image} {Denoising}},
	language = {en},
	booktitle = {Proceedings of the {IEEE}/{CVF} {Conference} on {Computer} {Vision} and {Pattern} {Recognition}},
	author = {Pang, Tongyao and Zheng, Huan and Quan, Yuhui and Ji, Hui},
	year = {2021},
	pages = {2043--2052},
}

@inproceedings{lucic_are_2018,
	title = {Are {GANs} {Created} {Equal}? {A} {Large}-{Scale} {Study}},
	booktitle = {Advances in {Neural} {Information} {Processing} {Systems}},
	author = {Lucic, Mario and Kurach, Karol and Michalski, Marcin and Gelly, Sylvain and Bousquet, Olivier},
	year = {2018},
}

@article{aggarwal_ensure_2023,
	title = {{ENSURE}: {A} {General} {Approach} for {Unsupervised} {Training} of {Deep} {Image} {Reconstruction} {Algorithms}},
	volume = {42},
	doi = {10.1109/TMI.2022.3224359},
	number = {4},
	journal = {IEEE Transactions on Medical Imaging},
	author = {Aggarwal, Hemant Kumar and Pramanik, Aniket and John, Maneesh and Jacob, Mathews},
	month = apr,
	year = {2023},
	pages = {1133--1144},
}

@article{aggarwal_modl_2019,
	title = {{MoDL}: {Model}-{Based} {Deep} {Learning} {Architecture} for {Inverse} {Problems}},
	volume = {38},
	doi = {10.1109/TMI.2018.2865356},
	number = {2},
	journal = {IEEE Transactions on Medical Imaging},
	author = {Aggarwal, Hemant K. and Mani, Merry P. and Jacob, Mathews},
	month = feb,
	year = {2019},
	pages = {394--405},
}

@inproceedings{aali_ambient_2024,
	title = {Ambient {Diffusion} {Posterior} {Sampling}: {Solving} {Inverse} {Problems} with {Diffusion} {Models} {Trained} on {Corrupted} {Data}},
	language = {en},
	booktitle = {International {Conference} on {Learning} {Representations}},
	author = {Aali, Asad and Daras, Giannis and Levac, Brett and Kumar, Sidharth and Dimakis, Alex and Tamir, Jon},
	month = oct,
	year = {2024},
}

@article{zou_dynamic_2021,
	title = {Dynamic {Imaging} {Using} a {Deep} {Generative} {SToRM} ({Gen}-{SToRM}) {Model}},
	volume = {40},
	doi = {10.1109/TMI.2021.3065948},
	number = {11},
	journal = {IEEE Transactions on Medical Imaging},
	author = {Zou, Qing and Ahmed, Abdul Haseeb and Nagpal, Prashant and Kruger, Stanley and Jacob, Mathews},
	month = nov,
	year = {2021},
	pages = {3102--3112},
}

@inproceedings{zhussip_training_2019,
	title = {Training {Deep} {Learning} {Based} {Image} {Denoisers} {From} {Undersampled} {Measurements} {Without} {Ground} {Truth} and {Without} {Image} {Prior}},
	doi = {10.1109/CVPR.2019.01050},
	booktitle = {2019 {IEEE}/{CVF} {Conference} on {Computer} {Vision} and {Pattern} {Recognition} ({CVPR})},
	author = {Zhussip, Magauiya and Soltanayev, Shakarim and Chun, Se Young},
	month = jun,
	year = {2019},
	pages = {10247--10256},
}

@article{zhou_dual-domain_2022,
	title = {Dual-domain self-supervised learning for accelerated non-{Cartesian} {MRI} reconstruction},
	volume = {81},
	doi = {10.1016/j.media.2022.102538},
	journal = {Medical Image Analysis},
	author = {Zhou, Bo and Schlemper, Jo and Dey, Neel and Mohseni Salehi, Seyed Sadegh and Sheth, Kevin and Liu, Chi and Duncan, James S. and Sofka, Michal},
	month = oct,
	year = {2022},
	pages = {102538},
}

@article{zhang_plug-and-play_2022,
	title = {Plug-and-{Play} {Image} {Restoration} {With} {Deep} {Denoiser} {Prior}},
	volume = {44},
	doi = {10.1109/TPAMI.2021.3088914},
	number = {10},
	journal = {IEEE Transactions on Pattern Analysis and Machine Intelligence},
	author = {Zhang, Kai and Li, Yawei and Zuo, Wangmeng and Zhang, Lei and Van Gool, Luc and Timofte, Radu},
	month = oct,
	year = {2022},
	pages = {6360--6376},
}

@inproceedings{zhang_cycle-consistent_2024,
	title = {Cycle-{Consistent} {Self}-{Supervised} {Learning} for {Improved} {Highly}-{Accelerated} {MRI} {Reconstruction}},
	doi = {10.1109/ISBI56570.2024.10635895},
	booktitle = {2024 {IEEE} {International} {Symposium} on {Biomedical} {Imaging} ({ISBI})},
	author = {Zhang, Chi and Demirel, Omer Burak and Akçakaya, Mehmet},
	month = may,
	year = {2024},
	pages = {1--5},
}

@article{zeng_review_2021,
	title = {A review on deep learning {MRI} reconstruction without fully sampled k-space},
	volume = {21},
	doi = {10.1186/s12880-021-00727-9},
	number = {1},
	journal = {BMC Medical Imaging},
	author = {Zeng, Gushan and Guo, Yi and Zhan, Jiaying and Wang, Zi and Lai, Zongying and Du, Xiaofeng and Qu, Xiaobo and Guo, Di},
	month = dec,
	year = {2021},
	pages = {195},
}

@inproceedings{yaman_zero-shot_2022,
	title = {Zero-{Shot} {Self}-{Supervised} {Learning} for {MRI} {Reconstruction}},
	language = {en},
	booktitle = {International {Conference} on {Learning} {Representations}},
	author = {Yaman, Burhaneddin and Hosseini, Seyed Amir Hossein and Akcakaya, Mehmet},
	month = jan,
	year = {2022},
}

@article{yaman_multi-mask_2022,
	title = {Multi-mask self-supervised learning for physics-guided neural networks in highly accelerated magnetic resonance imaging},
	volume = {35},
	doi = {10.1002/nbm.4798},
	language = {en},
	number = {12},
	journal = {NMR in Biomedicine},
	author = {Yaman, Burhaneddin and Gu, Hongyi and Hosseini, Seyed Amir Hossein and Demirel, Omer Burak and Moeller, Steen and Ellermann, Jutta and Uğurbil, Kâmil and Akçakaya, Mehmet},
	year = {2022},
	pages = {e4798},
}

@misc{xie_unsupervised_2020,
	title = {Unsupervised {Data} {Augmentation} for {Consistency} {Training}},
	doi = {10.48550/arXiv.1904.12848},
	publisher = {arXiv},
	author = {Xie, Qizhe and Dai, Zihang and Hovy, Eduard and Luong, Minh-Thang and Le, Quoc V.},
	month = nov,
	year = {2020},
	note = {arXiv:1904.12848 [cs]},
}

@inproceedings{xia_training_2019,
	title = {Training {Image} {Estimators} without {Image} {Ground} {Truth}},
	volume = {32},
	booktitle = {Advances in {Neural} {Information} {Processing} {Systems}},
	author = {Xia, Zhihao and Chakrabarti, Ayan},
	year = {2019},
}

@article{wang_parcel_2023,
	title = {{PARCEL}: {Physics}-{Based} {Unsupervised} {Contrastive} {Representation} {Learning} for {Multi}-{Coil} {MR} {Imaging}},
	volume = {20},
	doi = {10.1109/TCBB.2022.3213669},
	number = {5},
	journal = {IEEE/ACM Transactions on Computational Biology and Bioinformatics},
	author = {Wang, Shanshan and Wu, Ruoyou and Li, Cheng and Zou, Juan and Zhang, Ziyao and Liu, Qiegen and Xi, Yan and Zheng, Hairong},
	month = sep,
	year = {2023},
	pages = {2659--2670},
}

@misc{wang_k-band_2024,
	title = {K-band: {Self}-supervised {MRI} {Reconstruction} via {Stochastic} {Gradient} {Descent} over {K}-space {Subsets}},
	doi = {10.48550/arXiv.2308.02958},
	publisher = {arXiv},
	author = {Wang, Frederic and Qi, Han and Goyeneche, Alfredo De and Heckel, Reinhard and Lustig, Michael and Shimron, Efrat},
	month = may,
	year = {2024},
	note = {arXiv:2308.02958 [eess]},
}

@misc{wang_cmrxrecon_2023,
	title = {{CMRxRecon}: {An} open cardiac {MRI} dataset for the competition of accelerated image reconstruction},
	language = {en},
	journal = {arXiv.org},
	author = {Wang, Chengyan and Lyu, Jun and Wang, Shuo and Qin, Chen and Guo, Kunyuan and Zhang, Xinyu and Yu, Xiaotong and Li, Yan and Wang, Fanwen and Jin, Jianhua and Shi, Zhang and Xu, Ziqiang and Tian, Yapeng and Hua, Sha and Chen, Zhensen and Liu, Meng and Sun, Mengting and Kuang, Xutong and Wang, Kang and Wang, Haoran and Li, Hao and Chu, Yinghua and Yang, Guang and Bai, Wenjia and Zhuang, Xiahai and Wang, He and Qin, Jing and Qu, Xiaobo},
	month = sep,
	year = {2023},
}

@misc{wang_fully_2024,
	title = {Fully {Unsupervised} {Dynamic} {MRI} {Reconstruction} via {Diffeo}-{Temporal} {Equivariance}},
	doi = {10.48550/arXiv.2410.08646},
	publisher = {arXiv},
	author = {Wang, Andrew and Davies, Mike},
	month = oct,
	year = {2024},
	note = {arXiv:2410.08646 [eess]},
}

@inproceedings{wang_neural_2020,
	title = {Neural {Network}-{Based} {Reconstruction} in {Compressed} {Sensing} {MRI} {Without} {Fully}-{Sampled} {Training} {Data}},
	isbn = {978-3-030-61598-7},
	doi = {10.1007/978-3-030-61598-7_3},
	language = {en},
	booktitle = {Machine {Learning} for {Medical} {Image} {Reconstruction}},
	author = {Wang, Alan Q. and Dalca, Adrian V. and Sabuncu, Mert R.},
	editor = {Deeba, Farah and Johnson, Patricia and Würfl, Tobias and Ye, Jong Chul},
	year = {2020},
	pages = {27--37},
}

@inproceedings{ubaru_fast_2016,
	title = {Fast methods for estimating the {Numerical} rank of large matrices},
	language = {en},
	booktitle = {Proceedings of {The} 33rd {International} {Conference} on {Machine} {Learning}},
	publisher = {PMLR},
	author = {Ubaru, Shashanka and Saad, Yousef},
	month = jun,
	year = {2016},
	pages = {468--477},
}

@misc{tanabene_benchmarking_2024,
	title = {Benchmarking {3D} multi-coil {NC}-{PDNet} {MRI} reconstruction},
	doi = {10.48550/arXiv.2411.05883},
	publisher = {arXiv},
	author = {Tanabene, Asma and Radhakrishna, Chaithya Giliyar and Massire, Aurélien and Nadar, Mariappan S. and Ciuciu, Philippe},
	month = nov,
	year = {2024},
	note = {arXiv:2411.05883 [eess]},
}

@misc{tamir_unsupervised_2020,
	title = {Unsupervised {Deep} {Basis} {Pursuit}: {Learning} inverse problems without ground-truth data},
	doi = {10.48550/arXiv.1910.13110},
	publisher = {arXiv},
	author = {Tamir, Jonathan I. and Yu, Stella X. and Lustig, Michael},
	month = feb,
	year = {2020},
	note = {arXiv:1910.13110 [eess]},
}

@misc{tachella_unsure_2025,
	title = {{UNSURE}: self-supervised learning with {Unknown} {Noise} level and {Stein}'s {Unbiased} {Risk} {Estimate}},
	doi = {10.48550/arXiv.2409.01985},
	publisher = {arXiv},
	author = {Tachella, Julián and Davies, Mike and Jacques, Laurent},
	month = feb,
	year = {2025},
	note = {arXiv:2409.01985 [stat]},
}

@article{tachella_unsupervised_2022,
	title = {Unsupervised {Learning} {From} {Incomplete} {Measurements} for {Inverse} {Problems}},
	volume = {35},
	language = {en},
	journal = {Advances in Neural Information Processing Systems},
	author = {Tachella, Julián and Chen, Dongdong and Davies, Mike},
	month = dec,
	year = {2022},
	pages = {4983--4995},
}

@article{stein_estimation_1981,
	title = {Estimation of the {Mean} of a {Multivariate} {Normal} {Distribution}},
	volume = {9},
	doi = {10.1214/aos/1176345632},
	number = {6},
	journal = {The Annals of Statistics},
	author = {Stein, Charles M.},
	month = nov,
	year = {1981},
	pages = {1135--1151},
}

@inproceedings{sriram_end--end_2020,
	address = {Cham},
	title = {End-to-{End} {Variational} {Networks} for {Accelerated} {MRI} {Reconstruction}},
	isbn = {978-3-030-59713-9},
	doi = {10.1007/978-3-030-59713-9_7},
	language = {en},
	booktitle = {Medical {Image} {Computing} and {Computer} {Assisted} {Intervention} – {MICCAI} 2020},
	publisher = {Springer International Publishing},
	author = {Sriram, Anuroop and Zbontar, Jure and Murrell, Tullie and Defazio, Aaron and Zitnick, C. Lawrence and Yakubova, Nafissa and Knoll, Florian and Johnson, Patricia},
	editor = {Martel, Anne L. and Abolmaesumi, Purang and Stoyanov, Danail and Mateus, Diana and Zuluaga, Maria A. and Zhou, S. Kevin and Racoceanu, Daniel and Joskowicz, Leo},
	year = {2020},
	pages = {64--73},
}

@inproceedings{sidorov_deep_2019,
	title = {Deep {Hyperspectral} {Prior}: {Single}-{Image} {Denoising}, {Inpainting}, {Super}-{Resolution}},
	doi = {10.1109/ICCVW.2019.00477},
	booktitle = {2019 {IEEE}/{CVF} {International} {Conference} on {Computer} {Vision} {Workshop} ({ICCVW})},
	author = {Sidorov, Oleksii and Hardeberg, Jon Yngve},
	month = oct,
	year = {2019},
	pages = {3844--3851},
}

@article{shimron_implicit_2022,
	title = {Implicit data crimes: {Machine} learning bias arising from misuse of public data},
	volume = {119},
	doi = {10.1073/pnas.2117203119},
	number = {13},
	journal = {Proceedings of the National Academy of Sciences},
	author = {Shimron, Efrat and Tamir, Jonathan I. and Wang, Ke and Lustig, Michael},
	month = mar,
	year = {2022},
}

@article{shen_nerp_2024,
	title = {{NeRP}: {Implicit} {Neural} {Representation} {Learning} {With} {Prior} {Embedding} for {Sparsely} {Sampled} {Image} {Reconstruction}},
	volume = {35},
	doi = {10.1109/TNNLS.2022.3177134},
	number = {1},
	journal = {IEEE Transactions on Neural Networks and Learning Systems},
	author = {Shen, Liyue and Pauly, John and Xing, Lei},
	month = jan,
	year = {2024},
	pages = {770--782},
}

@article{shafique_mri_2024,
	title = {{MRI} {Recovery} with {Self}-{Calibrated} {Denoisers} without {Fully}-{Sampled} {Data}},
	volume = {38},
	doi = {10.1007/s10334-024-01207-1},
	number = {1},
	journal = {Magnetic Resonance Materials in Physics, Biology and Medicine},
	author = {Shafique, Muhammad and Liu, Sizhuo and Schniter, Philip and Ahmad, Rizwan},
	month = oct,
	year = {2024},
	pages = {53--66},
}

@inproceedings{senouf_self-supervised_2019,
	address = {Cham},
	title = {Self-supervised {Learning} of {Inverse} {Problem} {Solvers} in {Medical} {Imaging}},
	doi = {10.1007/978-3-030-33391-1_13},
	language = {en},
	booktitle = {Domain {Adaptation} and {Representation} {Transfer} and {Medical} {Image} {Learning} with {Less} {Labels} and {Imperfect} {Data}},
	publisher = {Springer International Publishing},
	author = {Senouf, Ortal and Vedula, Sanketh and Weiss, Tomer and Bronstein, Alex and Michailovich, Oleg and Zibulevsky, Michael},
	editor = {Wang, Qian and Milletari, Fausto and Nguyen, Hien V. and Albarqouni, Shadi and Cardoso, M. Jorge and Rieke, Nicola and Xu, Ziyue and Kamnitsas, Konstantinos and Patel, Vishal and Roysam, Badri and Jiang, Steve and Zhou, Kevin and Luu, Khoa and Le, Ngan},
	year = {2019},
	pages = {111--119},
}

@inproceedings{salimans_improved_2016,
	title = {Improved {Techniques} for {Training} {GANs}},
	volume = {29},
	booktitle = {Advances in {Neural} {Information} {Processing} {Systems}},
	publisher = {Curran Associates, Inc.},
	author = {Salimans, Tim and Goodfellow, Ian and Zaremba, Wojciech and Cheung, Vicki and Radford, Alec and Chen, Xi and Chen, Xi},
	year = {2016},
}

@article{ramani_monte-carlo_2008,
	title = {Monte-{Carlo} {Sure}: {A} {Black}-{Box} {Optimization} of {Regularization} {Parameters} for {General} {Denoising} {Algorithms}},
	volume = {17},
	doi = {10.1109/TIP.2008.2001404},
	number = {9},
	journal = {IEEE Transactions on Image Processing},
	author = {Ramani, Sathish and Blu, Thierry and Unser, Michael},
	month = sep,
	year = {2008},
	pages = {1540--1554},
}

@inproceedings{quan_dual-domain_2022,
	address = {Cham},
	title = {Dual-{Domain} {Self}-supervised {Learning} and {Model} {Adaption} for {Deep} {Compressive} {Imaging}},
	doi = {10.1007/978-3-031-20056-4_24},
	language = {en},
	booktitle = {Computer {Vision} – {ECCV} 2022},
	publisher = {Springer Nature Switzerland},
	author = {Quan, Yuhui and Qin, Xinran and Pang, Tongyao and Ji, Hui},
	editor = {Avidan, Shai and Brostow, Gabriel and Cissé, Moustapha and Farinella, Giovanni Maria and Hassner, Tal},
	year = {2022},
	pages = {409--426},
}

@article{pruessmann_sense_1999,
	title = {{SENSE}: {Sensitivity} encoding for fast {MRI}},
	volume = {42},
	doi = {10.1002/(SICI)1522-2594(199911)42:5<952::AID-MRM16>3.0.CO;2-S},
	language = {en},
	number = {5},
	journal = {Magnetic Resonance in Medicine},
	author = {Pruessmann, Klaas P. and Weiger, Markus and Scheidegger, Markus B. and Boesiger, Peter},
	year = {1999},
	pages = {952--962},
}

@inproceedings{pang_hir-diff_2024,
	title = {{HIR}-{Diff}: {Unsupervised} {Hyperspectral} {Image} {Restoration} {Via} {Improved} {Diffusion} {Models}},
	shorttitle = {{HIR}-{Diff}},
	language = {en},
	booktitle = {Proceedings of the {IEEE}/{CVF} {Conference} on {Computer} {Vision} and {Pattern} {Recognition}},
	author = {Pang, Li and Rui, Xiangyu and Cui, Long and Wang, Hongzhong and Meng, Deyu and Cao, Xiangyong},
	year = {2024},
	pages = {3005--3014},
}

@article{ongie_deep_2020,
	title = {Deep {Learning} {Techniques} for {Inverse} {Problems} in {Imaging}},
	volume = {1},
	doi = {10.1109/JSAIT.2020.2991563},
	number = {1},
	journal = {IEEE Journal on Selected Areas in Information Theory},
	author = {Ongie, Gregory and Jalal, Ajil and Metzler, Christopher A. and Baraniuk, Richard G. and Dimakis, Alexandros G. and Willett, Rebecca},
	month = may,
	year = {2020},
	pages = {39--56},
}

@article{ong_extreme_2020,
	title = {Extreme {MRI}: {Large}-scale volumetric dynamic imaging from continuous non-gated acquisitions},
	volume = {84},
	doi = {10.1002/mrm.28235},
	language = {en},
	number = {4},
	journal = {Magnetic Resonance in Medicine},
	author = {Ong, Frank and Zhu, Xucheng and Cheng, Joseph Y. and Johnson, Kevin M. and Larson, Peder E. Z. and Vasanawala, Shreyas S. and Lustig, Michael},
	year = {2020},
	pages = {1763--1780},
}

@inproceedings{ong_sigpy_2019,
	title = {{SigPy}: a python package for high performance iterative reconstruction},
	volume = {4819},
	booktitle = {Proceedings of the {ISMRM} 27th {Annual} {Meeting}, {Montreal}, {Quebec}, {Canada}},
	author = {Ong, Frank and Lustig, Michael},
	year = {2019},
}

@article{oh_unpaired_2020,
	title = {Unpaired {Deep} {Learning} for {Accelerated} {MRI} {Using} {Optimal} {Transport} {Driven} {CycleGAN}},
	volume = {6},
	doi = {10.1109/TCI.2020.3018562},
	journal = {IEEE Transactions on Computational Imaging},
	author = {Oh, Gyutaek and Sim, Byeongsu and Chung, HyungJin and Sunwoo, Leonard and Ye, Jong Chul},
	year = {2020},
	pages = {1285--1296},
}

@inproceedings{mukherjee_end--end_2021,
	title = {End-to-end reconstruction meets data-driven regularization for inverse problems},
	volume = {34},
	booktitle = {Advances in {Neural} {Information} {Processing} {Systems}},
	author = {Mukherjee, Subhadip and Carioni, Marcello and Öktem, Ozan and Schönlieb, Carola-Bibiane},
	year = {2021},
	pages = {21413--21425},
}

@misc{moran_noisier2noise_2019,
	title = {{Noisier2Noise}: {Learning} to {Denoise} from {Unpaired} {Noisy} {Data}},
	doi = {10.48550/arXiv.1910.11908},
	publisher = {arXiv},
	author = {Moran, Nick and Schmidt, Dan and Zhong, Yu and Coady, Patrick},
	month = oct,
	year = {2019},
	note = {arXiv:1910.11908 [cs, eess]},
}

@misc{millard_clean_2024,
	title = {Clean self-supervised {MRI} reconstruction from noisy, sub-sampled training data with {Robust} {SSDU}},
	doi = {10.48550/arXiv.2210.01696},
	publisher = {arXiv},
	author = {Millard, Charles and Chiew, Mark},
	month = jun,
	year = {2024},
	note = {arXiv:2210.01696 [eess]},
}

@article{millard_theoretical_2023,
	title = {A {Theoretical} {Framework} for {Self}-{Supervised} {MR} {Image} {Reconstruction} {Using} {Sub}-{Sampling} via {Variable} {Density} {Noisier2Noise}},
	volume = {9},
	doi = {10.1109/TCI.2023.3299212},
	journal = {IEEE Transactions on Computational Imaging},
	author = {Millard, Charles and Chiew, Mark},
	year = {2023},
	pages = {707--720},
}

@misc{metzler_unsupervised_2020,
	title = {Unsupervised {Learning} with {Stein}'s {Unbiased} {Risk} {Estimator}},
	doi = {10.48550/arXiv.1805.10531},
	publisher = {arXiv},
	author = {Metzler, Christopher A. and Mousavi, Ali and Heckel, Reinhard and Baraniuk, Richard G.},
	month = jul,
	year = {2020},
	note = {arXiv:1805.10531 [stat]},
}

@article{lyu_m4raw_2023,
	title = {{M4Raw}: {A} multi-contrast, multi-repetition, multi-channel {MRI} k-space dataset for low-field {MRI} research},
	volume = {10},
	doi = {10.1038/s41597-023-02181-4},
	language = {en},
	number = {1},
	journal = {Scientific Data},
	author = {Lyu, Mengye and Mei, Lifeng and Huang, Shoujin and Liu, Sixing and Li, Yi and Yang, Kexin and Liu, Yilong and Dong, Yu and Dong, Linzheng and Wu, Ed X.},
	month = may,
	year = {2023},
	pages = {264},
}

@article{lustig_compressed_2008,
	title = {Compressed {Sensing} {MRI}},
	volume = {25},
	doi = {10.1109/MSP.2007.914728},
	number = {2},
	journal = {IEEE Signal Processing Magazine},
	author = {Lustig, Michael and Donoho, David L. and Santos, Juan M. and Pauly, John M.},
	month = mar,
	year = {2008},
	pages = {72--82},
}

@misc{luo_unsupervised_2025,
	title = {Unsupervised {Accelerated} {MRI} {Reconstruction} via {Ground}-{Truth}-{Free} {Flow} {Matching}},
	doi = {10.48550/arXiv.2502.17174},
	publisher = {arXiv},
	author = {Luo, Xinzhe and Li, Yingzhen and Qin, Chen},
	month = feb,
	year = {2025},
	note = {arXiv:2502.17174 [eess]},
}

@article{liu_rare_2020,
	title = {{RARE}: {Image} {Reconstruction} {Using} {Deep} {Priors} {Learned} {Without} {Groundtruth}},
	volume = {14},
	doi = {10.1109/JSTSP.2020.2998402},
	number = {6},
	journal = {IEEE Journal of Selected Topics in Signal Processing},
	author = {Liu, Jiaming and Sun, Yu and Eldeniz, Cihat and Gan, Weijie and An, Hongyu and Kamilov, Ulugbek S.},
	month = oct,
	year = {2020},
	pages = {1088--1099},
}

@article{li_progressive_2024,
	title = {Progressive dual-domain-transfer {cycleGAN} for unsupervised {MRI} reconstruction},
	volume = {563},
	doi = {10.1016/j.neucom.2023.126934},
	journal = {Neurocomputing},
	author = {Li, Bowen and Wang, Zhiwen and Yang, Ziyuan and Xia, Wenjun and Zhang, Yi},
	month = jan,
	year = {2024},
	pages = {126934},
}

@article{lei_wasserstein_2021,
	title = {Wasserstein {GANs} for {MR} {Imaging}: {From} {Paired} to {Unpaired} {Training}},
	volume = {40},
	doi = {10.1109/TMI.2020.3022968},
	number = {1},
	journal = {IEEE Transactions on Medical Imaging},
	author = {Lei, Ke and Mardani, Morteza and Pauly, John M. and Vasanawala, Shreyas S.},
	month = jan,
	year = {2021},
	pages = {105--115},
}

@article{korkmaz_unsupervised_2022,
	title = {Unsupervised {MRI} {Reconstruction} via {Zero}-{Shot} {Learned} {Adversarial} {Transformers}},
	volume = {41},
	doi = {10.1109/TMI.2022.3147426},
	number = {7},
	journal = {IEEE Transactions on Medical Imaging},
	author = {Korkmaz, Yilmaz and Dar, Salman U. H. and Yurt, Mahmut and Özbey, Muzaffer and Çukur, Tolga},
	month = jul,
	year = {2022},
	pages = {1747--1763},
}

@article{klug_analyzing_2023,
	title = {Analyzing the {Sample} {Complexity} of {Self}-{Supervised} {Image} {Reconstruction} {Methods}},
	volume = {36},
	language = {en},
	journal = {Advances in Neural Information Processing Systems},
	author = {Klug, Tobit and Atik, Dogukan and Heckel, Reinhard},
	month = dec,
	year = {2023},
	pages = {65869--65893},
}

@misc{kawar_gsure-based_2024,
	title = {{GSURE}-{Based} {Diffusion} {Model} {Training} with {Corrupted} {Data}},
	doi = {10.48550/arXiv.2305.13128},
	publisher = {arXiv},
	author = {Kawar, Bahjat and Elata, Noam and Michaeli, Tomer and Elad, Michael},
	month = jun,
	year = {2024},
	note = {arXiv:2305.13128 [eess]},
}

@misc{janjusevic_self-supervised_2025,
	title = {Self-{Supervised} {Noise} {Adaptive} {MRI} {Denoising} via {Repetition} to {Repetition} ({Rep2Rep}) {Learning}},
	doi = {10.48550/arXiv.2504.17698},
	publisher = {arXiv},
	author = {Janjušević, Nikola and Chen, Jingjia and Ginocchio, Luke and Bruno, Mary and Huang, Yuhui and Wang, Yao and Chandarana, Hersh and Feng, Li},
	month = apr,
	year = {2025},
	note = {arXiv:2504.17698 [eess]},
}

@article{huang_self-supervised_2024,
	title = {Self-{Supervised} {Deep} {Unrolled} {Reconstruction} {Using} {Regularization} by {Denoising}},
	volume = {43},
	doi = {10.1109/TMI.2023.3332614},
	number = {3},
	journal = {IEEE Transactions on Medical Imaging},
	author = {Huang, Peizhou and Zhang, Chaoyi and Zhang, Xiaoliang and Li, Xiaojuan and Dong, Liang and Ying, Leslie},
	month = mar,
	year = {2024},
	pages = {1203--1213},
}

@inproceedings{huang_deep_2019,
	title = {Deep {MRI} {Reconstruction} without {Ground} {Truth} for {Training}},
	booktitle = {Proc. {Intl}. {Soc}. {Mag}. {Reson}. {Med}. 27},
	author = {Huang, Peizhou and Zhang, Chaoyi and Li, Hongyu and Gaire, Sunil Kumar and Liu, Ruiying and Zhang, Xiaoliang and Li, Xiaojuan and Ying, Leslie},
	year = {2019},
}

@misc{hu_spicer_2024,
	title = {{SPICER}: {Self}-{Supervised} {Learning} for {MRI} with {Automatic} {Coil} {Sensitivity} {Estimation} and {Reconstruction}},
	doi = {10.48550/arXiv.2210.02584},
	publisher = {arXiv},
	author = {Hu, Yuyang and Gan, Weijie and Ying, Chunwei and Wang, Tongyao and Eldeniz, Cihat and Liu, Jiaming and Chen, Yasheng and An, Hongyu and Kamilov, Ulugbek S.},
	month = jun,
	year = {2024},
	note = {arXiv:2210.02584 [eess]},
}

@inproceedings{hu_self-supervised_2021,
	title = {Self-supervised {Learning} for {MRI} {Reconstruction} with a {Parallel} {Network} {Training} {Framework}},
	doi = {10.1007/978-3-030-87231-1_37},
	language = {en},
	booktitle = {Medical {Image} {Computing} and {Computer} {Assisted} {Intervention} – {MICCAI} 2021},
	author = {Hu, Chen and Li, Cheng and Wang, Haifeng and Liu, Qiegen and Zheng, Hairong and Wang, Shanshan},
	editor = {de Bruijne, Marleen and Cattin, Philippe C. and Cotin, Stéphane and Padoy, Nicolas and Speidel, Stefanie and Zheng, Yefeng and Essert, Caroline},
	year = {2021},
	pages = {382--391},
}

@article{heckel_deep_2024,
	title = {Deep learning for accelerated and robust {MRI} reconstruction},
	volume = {37},
	doi = {10.1007/s10334-024-01173-8},
	language = {en},
	number = {3},
	journal = {Magnetic Resonance Materials in Physics, Biology and Medicine},
	author = {Heckel, Reinhard and Jacob, Mathews and Chaudhari, Akshay and Perlman, Or and Shimron, Efrat},
	month = jul,
	year = {2024},
	pages = {335--368},
}

@article{he_comparative_2020,
	title = {A {Comparative} {Study} of {Unsupervised} {Deep} {Learning} {Methods} for {MRI} {Reconstruction}},
	language = {en},
	journal = {Investigative Magnetic Resonance Imaging},
	author = {He, Zhuonan and Quan, Cong and Wang, Siyuan and Zhu, Yuanzheng and Zhang, Minghui and Zhu, Yanjie and Liu, Qiegen and He, Zhuonan and Quan, Cong and Wang, Siyuan and Zhu, Yuanzheng and Zhang, Minghui and Zhu, Yanjie and Liu, Qiegen},
	year = {2020},
	pages = {179--195},
}

@article{hammernik_learning_2018,
	title = {Learning a variational network for reconstruction of accelerated {MRI} data},
	volume = {79},
	doi = {10.1002/mrm.26977},
	language = {en},
	number = {6},
	journal = {Magnetic Resonance in Medicine},
	author = {Hammernik, Kerstin and Klatzer, Teresa and Kobler, Erich and Recht, Michael P. and Sodickson, Daniel K. and Pock, Thomas and Knoll, Florian},
	year = {2018},
	pages = {3055--3071},
}

@inproceedings{gan_deep_2021,
	title = {Deep {Image} {Reconstruction} {Using} {Unregistered} {Measurements} {Without} {Groundtruth}},
	doi = {10.1109/ISBI48211.2021.9434079},
	booktitle = {2021 {IEEE} 18th {International} {Symposium} on {Biomedical} {Imaging} ({ISBI})},
	author = {Gan, Weijie and Sun, Yu and Eldeniz, Cihat and Liu, Jiaming and An, Hongyu and Kamilov, Ulugbek S.},
	month = apr,
	year = {2021},
	pages = {1531--1534},
}

@article{freifeld_transformations_2017,
	title = {Transformations {Based} on {Continuous} {Piecewise}-{Affine} {Velocity} {Fields}},
	volume = {39},
	doi = {10.1109/TPAMI.2016.2646685},
	number = {12},
	journal = {IEEE Transactions on Pattern Analysis and Machine Intelligence},
	author = {Freifeld, Oren and Hauberg, Søren and Batmanghelich, Kayhan and Fisher, Jonn W.},
	month = dec,
	year = {2017},
	pages = {2496--2509},
}

@article{ericsson_self-supervised_2022,
	title = {Self-{Supervised} {Representation} {Learning}: {Introduction}, advances, and challenges},
	volume = {39},
	doi = {10.1109/MSP.2021.3134634},
	number = {3},
	journal = {IEEE Signal Processing Magazine},
	author = {Ericsson, Linus and Gouk, Henry and Loy, Chen Change and Hospedales, Timothy M.},
	month = may,
	year = {2022},
	pages = {42--62},
}

@article{eldar_generalized_2009,
	title = {Generalized {SURE} for {Exponential} {Families}: {Applications} to {Regularization}},
	volume = {57},
	doi = {10.1109/TSP.2008.2008212},
	number = {2},
	journal = {IEEE Transactions on Signal Processing},
	author = {Eldar, Yonina C.},
	month = feb,
	year = {2009},
	pages = {471--481},
}

@misc{desai_skm-tea_2022,
	title = {{SKM}-{TEA}: {A} {Dataset} for {Accelerated} {MRI} {Reconstruction} with {Dense} {Image} {Labels} for {Quantitative} {Clinical} {Evaluation}},
	doi = {10.48550/arXiv.2203.06823},
	publisher = {arXiv},
	author = {Desai, Arjun D. and Schmidt, Andrew M. and Rubin, Elka B. and Sandino, Christopher M. and Black, Marianne S. and Mazzoli, Valentina and Stevens, Kathryn J. and Boutin, Robert and Ré, Christopher and Gold, Garry E. and Hargreaves, Brian A. and Chaudhari, Akshay S.},
	month = mar,
	year = {2022},
	note = {arXiv:2203.06823 [eess]},
}

@misc{desai_noise2recon_2022,
	title = {{Noise2Recon}: {Enabling} {Joint} {MRI} {Reconstruction} and {Denoising} with {Semi}-{Supervised} and {Self}-{Supervised} {Learning}},
	doi = {10.48550/arXiv.2110.00075},
	publisher = {arXiv},
	author = {Desai, Arjun D. and Ozturkler, Batu M. and Sandino, Christopher M. and Boutin, Robert and Willis, Marc and Vasanawala, Shreyas and Hargreaves, Brian A. and Ré, Christopher M. and Pauly, John M. and Chaudhari, Akshay S.},
	month = oct,
	year = {2022},
	note = {arXiv:2110.00075 [eess]},
}

@misc{desai_vortex_2022,
	title = {{VORTEX}: {Physics}-{Driven} {Data} {Augmentations} {Using} {Consistency} {Training} for {Robust} {Accelerated} {MRI} {Reconstruction}},
	doi = {10.48550/arXiv.2111.02549},
	publisher = {arXiv},
	author = {Desai, Arjun D. and Gunel, Beliz and Ozturkler, Batu M. and Beg, Harris and Vasanawala, Shreyas and Hargreaves, Brian A. and Ré, Christopher and Pauly, John M. and Chaudhari, Akshay S.},
	month = jun,
	year = {2022},
	note = {arXiv:2111.02549 [eess]},
}

@inproceedings{darestani_test-time_2022,
	title = {Test-{Time} {Training} {Can} {Close} the {Natural} {Distribution} {Shift} {Performance} {Gap} in {Deep} {Learning} {Based} {Compressed} {Sensing}},
	booktitle = {Proceedings of the 39th {International} {Conference} on {Machine} {Learning}},
	publisher = {PMLR},
	author = {Darestani, Mohammad Zalbagi and Liu, Jiayu and Heckel, Reinhard},
	month = jun,
	year = {2022},
	pages = {4754--4776},
}

@misc{darestani_accelerated_2021,
	title = {Accelerated {MRI} with {Un}-trained {Neural} {Networks}},
	doi = {10.48550/arXiv.2007.02471},
	publisher = {arXiv},
	author = {Darestani, Mohammad Zalbagi and Heckel, Reinhard},
	month = apr,
	year = {2021},
	note = {arXiv:2007.02471 [eess]},
}

@misc{daras_ambient_2023,
	title = {Ambient {Diffusion}: {Learning} {Clean} {Distributions} from {Corrupted} {Data}},
	doi = {10.48550/arXiv.2305.19256},
	publisher = {arXiv},
	author = {Daras, Giannis and Shah, Kulin and Dagan, Yuval and Gollakota, Aravind and Dimakis, Alexandros G. and Klivans, Adam},
	month = may,
	year = {2023},
	note = {arXiv:2305.19256 [cs, math]},
}

@misc{daras_survey_2024,
	title = {A {Survey} on {Diffusion} {Models} for {Inverse} {Problems}},
	doi = {10.48550/arXiv.2410.00083},
	publisher = {arXiv},
	author = {Daras, Giannis and Chung, Hyungjin and Lai, Chieh-Hsin and Mitsufuji, Yuki and Ye, Jong Chul and Milanfar, Peyman and Dimakis, Alexandros G. and Delbracio, Mauricio},
	month = sep,
	year = {2024},
	note = {arXiv:2410.00083 [cs]},
}

@inproceedings{cole_fast_2021,
	title = {Fast {Unsupervised} {MRI} {Reconstruction} {Without} {Fully}-{Sampled} {Ground} {Truth} {Data} {Using} {Generative} {Adversarial} {Networks}},
	doi = {10.1109/ICCVW54120.2021.00444},
	booktitle = {2021 {IEEE}/{CVF} {International} {Conference} on {Computer} {Vision} {Workshops} ({ICCVW})},
	author = {Cole, Elizabeth K. and Ong, Frank and Vasanawala, Shreyas S. and Pauly, John M.},
	month = oct,
	year = {2021},
	pages = {3971--3980},
}

@article{chung_score-based_2022,
	title = {Score-based diffusion models for accelerated {MRI}},
	volume = {80},
	doi = {10.1016/j.media.2022.102479},
	urldate = {2025-04-07},
	journal = {Medical Image Analysis},
	author = {Chung, Hyungjin and Ye, Jong Chul},
	month = aug,
	year = {2022},
	pages = {102479},
}

@article{chen_ai-based_2022,
	title = {{AI}-{Based} {Reconstruction} for {Fast} {MRI}—{A} {Systematic} {Review} and {Meta}-{Analysis}},
	volume = {110},
	doi = {10.1109/JPROC.2022.3141367},
	number = {2},
	urldate = {2025-04-04},
	journal = {Proceedings of the IEEE},
	author = {Chen, Yutong and Schönlieb, Carola-Bibiane and Liò, Pietro and Leiner, Tim and Dragotti, Pier Luigi and Wang, Ge and Rueckert, Daniel and Firmin, David and Yang, Guang},
	month = feb,
	year = {2022},
	pages = {224--245},
}

@article{chen_imaging_2023,
	title = {Imaging {With} {Equivariant} {Deep} {Learning}: {From} unrolled network design to fully unsupervised learning},
	volume = {40},
	doi = {10.1109/MSP.2022.3205430},
	number = {1},
	journal = {IEEE Signal Processing Magazine},
	author = {Chen, Dongdong and Davies, Mike and Ehrhardt, Matthias J. and Schönlieb, Carola-Bibiane and Sherry, Ferdia and Tachella, Julián},
	month = jan,
	year = {2023},
	pages = {134--147},
}

@article{chaudhari_prospective_2021,
	title = {Prospective {Deployment} of {Deep} {Learning} in {MRI}: {A} {Framework} for {Important} {Considerations}, {Challenges}, and {Recommendations} for {Best} {Practices}},
	volume = {54},
	doi = {10.1002/jmri.27331},
	language = {en},
	number = {2},
	urldate = {2025-04-25},
	journal = {Journal of Magnetic Resonance Imaging},
	author = {Chaudhari, Akshay S. and Sandino, Christopher M. and Cole, Elizabeth K. and Larson, David B. and Gold, Garry E. and Vasanawala, Shreyas S. and Lungren, Matthew P. and Hargreaves, Brian A. and Langlotz, Curtis P.},
	year = {2021},
	pages = {357--371},
}

@article{breger_study_2025,
	title = {A study of why we need to reassess full reference image quality assessment with medical images},
	doi = {10.1007/s10278-025-01462-1},
	journal = {Journal of Imaging Informatics in Medicine},
	author = {Breger, Anna and Biguri, Ander and Landman, Malena Sabaté and Selby, Ian and Amberg, Nicole and Brunner, Elisabeth and Gröhl, Janek and Hatamikia, Sepideh and Karner, Clemens and Ning, Lipeng and Dittmer, Sören and Roberts, Michael and Collaboration, AIX-COVNET and Schönlieb, Carola-Bibiane},
	month = mar,
	year = {2025},
}

@inproceedings{bodrito_trainable_2021,
	title = {A {Trainable} {Spectral}-{Spatial} {Sparse} {Coding} {Model} for {Hyperspectral} {Image} {Restoration}},
	volume = {34},
	urldate = {2025-05-12},
	booktitle = {Advances in {Neural} {Information} {Processing} {Systems}},
	author = {Bodrito, Theo and Zouaoui, Alexandre and Chanussot, Jocelyn and Mairal, Julien},
	year = {2021},
	pages = {5430--5442},
}

@inproceedings{alcalar_convex_2024,
	title = {A {Convex} {Compressibility}-{Inspired} {Unsupervised} {Loss} {Function} for {Physics}-{Driven} {Deep} {Learning} {Reconstruction}},
	doi = {10.1109/ISBI56570.2024.10635138},
	booktitle = {2024 {IEEE} {International} {Symposium} on {Biomedical} {Imaging} ({ISBI})},
	author = {Alçalar, Yaşar Utku and Gülle, Merve and Akçakaya, Mehmet},
	month = may,
	year = {2024},
	pages = {1--5},
}

@article{akcakaya_unsupervised_2022,
	title = {Unsupervised {Deep} {Learning} {Methods} for {Biological} {Image} {Reconstruction} and {Enhancement}: {An} overview from a signal processing perspective},
	volume = {39},
	doi = {10.1109/MSP.2021.3119273},
	number = {2},
	journal = {IEEE Signal Processing Magazine},
	author = {Akçakaya, Mehmet and Yaman, Burhaneddin and Chung, Hyungjin and Ye, Jong Chul},
	month = mar,
	year = {2022},
	pages = {28--44},
}

@article{aali_gsure_2024,
	title = {{GSURE} {Denoising} enables training of higher quality generative priors for accelerated {Multi}-{Coil} {MRI} {Reconstruction}},
	journal = {ISMRM 2024},
	author = {Aali, Asad and Arvinte, Marius and Kumar, Sidharth and Arefeen, Yamin I. and Tamir, Jonathan I.},
	month = apr,
	year = {2024},
}

@inproceedings{acar_self-supervised_2021,
	title = {Self-supervised {Dynamic} {MRI} {Reconstruction}},
	doi = {10.1007/978-3-030-88552-6_4},
	booktitle = {Machine {Learning} for {Medical} {Image} {Reconstruction}},
	publisher = {Springer International Publishing},
	author = {Acar, Mert and Çukur, Tolga and Öksüz, Ilkay},
	year = {2021},
	pages = {35--44},
}

@inproceedings{chen_robust_2022,
	title = {Robust {Equivariant} {Imaging}: a fully unsupervised framework for learning to image from noisy and partial measurements},
	doi = {10.1109/CVPR52688.2022.00556},
	booktitle = {2022 {IEEE}/{CVF} {Conference} on {Computer} {Vision} and {Pattern} {Recognition} ({CVPR})},
	author = {Chen, Dongdong and Tachella, Julián and Davies, Mike E.},
	month = jun,
	year = {2022},
}

@article{gan_deformation-compensated_2022,
	title = {Deformation-{Compensated} {Learning} for {Image} {Reconstruction} {Without} {Ground} {Truth}},
	volume = {41},
	issn = {1558-254X},
	doi = {10.1109/TMI.2022.3163018},
	number = {9},
	journal = {IEEE Transactions on Medical Imaging},
	author = {Gan, Weijie and Sun, Yu and Eldeniz, Cihat and Liu, Jiaming and An, Hongyu and Kamilov, Ulugbek S.},
	month = sep,
	year = {2022},
	pages = {2371--2384},
}

@article{yaman_self-supervised_2020,
	title = {Self-supervised learning of physics-guided reconstruction neural networks without fully sampled reference data},
	volume = {84},
	doi = {10.1002/mrm.28378},
	language = {en},
	number = {6},
	journal = {Magnetic Resonance in Medicine},
	author = {Yaman, Burhaneddin and Hosseini, Seyed Amir Hossein and Moeller, Steen and Ellermann, Jutta and Uğurbil, Kâmil and Akçakaya, Mehmet},
	year = {2020},
	pages = {3172--3191},
}

@inproceedings{ronneberger_u-net_2015,
	title = {U-{Net}: {Convolutional} {Networks} for {Biomedical} {Image} {Segmentation}},
	doi = {10.1007/978-3-319-24574-4_28},
	booktitle = {Medical {Image} {Computing} and {Computer}-{Assisted} {Intervention} – {MICCAI} 2015},
	author = {Ronneberger, Olaf and Fischer, Philipp and Brox, Thomas},
	editor = {Navab, Nassir and Hornegger, Joachim and Wells, William M. and Frangi, Alejandro F.},
	year = {2015},
	pages = {234--241},
}

@article{qin_convolutional_2019,
	title = {Convolutional {Recurrent} {Neural} {Networks} for {Dynamic} {MR} {Image} {Reconstruction}},
	volume = {38},
	doi = {10.1109/TMI.2018.2863670},
	number = {1},
	journal = {IEEE Transactions on Medical Imaging},
	author = {Qin, Chen and Schlemper, Jo and Caballero, Jose and Price, Anthony N. and Hajnal, Joseph V. and Rueckert, Daniel},
	month = jan,
	year = {2019},
	pages = {280--290},
}

@inproceedings{pajot_unsupervised_2018,
	title = {Unsupervised {Adversarial} {Image} {Reconstruction}},
	booktitle = {International {Conference} on {Learning} {Representations}},
	author = {Pajot, Arthur and Bezenac, Emmanuel de and Gallinari, Patrick},
	month = sep,
	year = {2018},
}

@article{meng_large-scale_2021,
	title = {A {Large}-{Scale} {Benchmark} {Data} {Set} for {Evaluating} {Pansharpening} {Performance}: {Overview} and {Implementation}},
	doi = {10.1109/MGRS.2020.2976696},
	journal = {IEEE Geoscience and Remote Sensing Magazine},
	author = {Meng, Xiangchao and Xiong, Yiming and Shao, Feng and Shen, Huanfeng and Sun, Weiwei and Yang, Gang and Yuan, Qiangqiang and Fu, Randi and Zhang, Hongyan},
	month = mar,
	year = {2021},
}

@article{hendriksen_noise2inverse_2020,
	title = {{Noise2Inverse}: {Self}-{Supervised} {Deep} {Convolutional} {Denoising} for {Tomography}},
	volume = {6},
	doi = {10.1109/TCI.2020.3019647},
	journal = {IEEE Transactions on Computational Imaging},
	author = {Hendriksen, Allard Adriaan and Pelt, Daniël Maria and Batenburg, K. Joost},
	year = {2020},
	pages = {1320--1335},
}

@inproceedings{fabian_data_2021,
	title = {Data augmentation for deep learning based accelerated {MRI} reconstruction with limited data},
	urldate = {2024-02-14},
	booktitle = {Proceedings of the 38th {International} {Conference} on {Machine} {Learning}},
	author = {Fabian, Zalan and Heckel, Reinhard and Soltanolkotabi, Mahdi},
	month = jul,
	year = {2021},
	pages = {3057--3067},
}

@inproceedings{chen_equivariant_2021,
	title = {Equivariant {Imaging}: {Learning} {Beyond} the {Range} {Space}},
	doi = {10.1109/ICCV48922.2021.00434},
	language = {English},
	booktitle = {2021 {IEEE}/{CVF} {International} {Conference} on {Computer} {Vision} ({ICCV})},
	author = {Chen, Dongdong and Tachella, Julián and Davies, Mike E.},
	month = oct,
	year = {2021},
}

@inproceedings{bora_ambientgan_2018,
	title = {{AmbientGAN}: {Generative} models from lossy measurements},
	booktitle = {International {Conference} on {Learning} {Representations}},
	author = {Bora, Ashish and Price, Eric and Dimakis, Alexandros G.},
	month = feb,
	year = {2018},
}
